\documentclass[nonacm]{acmart}

\usepackage{amsmath}
\usepackage{amsfonts}
\usepackage{algorithmic}
\usepackage{threeparttable}
\usepackage{graphicx}
\usepackage{adjustbox}
\usepackage{textcomp}
\usepackage{xcolor}
\usepackage{multirow}
\usepackage{float}
\restylefloat{table}
\usepackage{url}

\def\BibTeX{{\rm B\kern-.05em{\sc i\kern-.025em b}\kern-.08em
    T\kern-.1667em\lower.7ex\hbox{E}\kern-.125emX}}
    
\usepackage{lipsum}                     %
\usepackage{xargs}  
\usepackage{color,soul}%

\usepackage{blindtext}

\usepackage[caption=false]{subfig}

\usepackage{outlines}
\usepackage{booktabs}
\usepackage{tabularx}

\usepackage{tcolorbox}
\tcbset{mybox/.style={colback=white, colframe=blue, left=1mm, right=1mm, 
  fonttitle=\bfseries}, fontupper=\small,
  before upper=\setlength{\parindent}{1em}\everypar{{\setbox0\lastbox}\everypar{}},
}
\newtcolorbox{mybox}[1][]{mybox,#1}

\newcommand{\science}{\begingroup
\setbox0=\hbox{\includegraphics[width=3.5mm, height=3.5mm]{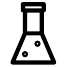}}\parbox{\wd0}{\box0}\endgroup}

\newcommand{\oss}{\begingroup
\setbox0=\hbox{\includegraphics[width=3.5mm, height=3.5mm]{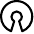}}\parbox{\wd0}{\box0}\endgroup}

\newcommand{\szmodifyok}[2]{#2}

\usepackage{wrapfig}

\def\projectname{\emph{Astropy}}
\def\numsurveyee{\emph{80}}
\def\numCRGSubgraph{697}

\def\repoa{\emph{poliastro}}
\def\repob{\emph{sbpy}}
\def\repoc{\emph{astroquery}}

\def\RQone{\emph{What are the major challenges when interdisciplinary experts collaborate to develop and maintain the Astropy project?}}

\def\RQtwo{\emph{What are the main factors for sustaining the Astropy project community?}}

\def\RQthree{\emph{What are the intentions of cross-project collaboration and corresponding challenges in the Astropy project?}}

\graphicspath{ {./Figures/} }  

\usepackage{enumitem}
\setitemize{noitemsep,topsep=0pt,parsep=0pt,partopsep=0pt,leftmargin=20pt}

\usepackage{titlesec}
\setstcolor{red}

\titlespacing*{\section}
{0pt}{1ex plus 1ex minus .1ex}{1ex plus .2ex}
\titlespacing*{\subsection}
{0pt}{1ex plus 1ex minus .1ex}{1ex plus .2ex}

\setcopyright{none}
\renewcommand\footnotetextcopyrightpermission[1]{} %
\AtBeginDocument{%
  \providecommand\BibTeX{{%
    \normalfont B\kern-0.5em{\scshape i\kern-0.25em b}\kern-0.8em\TeX}}}

\acmPrice{15.00}
\acmISBN{978-1-4503-XXXX-X/18/06}

\begin{document}

\title{How to Sustain a Scientific Open-Source Software Ecosystem: Learning from the Astropy Project}

\author{Jiayi Sun}
\affiliation{%
  \institution{University of Toronto}
  \country{Canada}}
\email{jysun@ece.utoronto.ca}

\author{Aarya Patil}
\affiliation{%
  \institution{University of Toronto}
  \country{Canada}}
\email{aarya.patil@mail.utoronto.ca}

\author{Youhai Li}
\affiliation{%
  \institution{University of Toronto}
  \country{Canada}}
\email{youhai.li@mail.utoronto.ca}

\author{Jin L.C. Guo}
\affiliation{%
  \institution{McGill University}
  \country{Canada}}
\email{jguo@cs.mcgill.ca}

\author{Shurui Zhou}
\affiliation{%
  \institution{University of Toronto}
  \country{Canada}}
\email{shurui.zhou@utoronto.ca}

\renewcommand{\shortauthors}{Sun et al.}

\begin{abstract}
Scientific open-source software (OSS) has greatly benefited research communities through its transparent and collaborative nature. 
Given its critical role in scientific research, ensuring the sustainability of such software has become vital. 
Earlier studies have proposed sustainability strategies for conventional scientific software and open-source communities. 
However, it remains unclear whether these solutions can be easily adapted to the integrated framework of scientific OSS and its larger ecosystem. 
This study examines the challenges and opportunities to enhance the sustainability of scientific OSS in the context of -- \emph{interdisciplinary collaboration}, 
\emph{open-source community}, and \emph{multi-project ecosystem}. 
We conducted a case study on a widely-used software ecosystem in the astrophysics domain, the Astropy Project, using a mixed-methods design approach. This approach includes an interview with core contributors regarding their participation in an interdisciplinary team, a survey of disengaged contributors about their motivations for contribution, reasons for disengagement, and suggestions for sustaining the communities, and finally, an analysis of cross-referenced issues and pull requests to understand best practices for collaboration on the ecosystem level. 
Our study reveals the implications of major challenges for sustaining scientific OSS and proposes concrete suggestions for tackling these challenges. 
\end{abstract}

\maketitle
\section{Introduction}\label{sec-intro}
Scientific software\footnote{It is also called science software, research software, or academic software. In this paper, we employ the phrase ``scientific software'' to maintain consistency.} development refers to the process 
of building software that is used in scientific disciplines such as biology, physics, geoscience, chemistry, and astronomy~\cite{arvanitou_software_2021,gruenpeter_2021,wiki_sci_sw,lamprecht2020towards}.
Given the critical role software plays in data processing and computation, scientific software has become the foundation for new discoveries~\cite{segal2004professional,ko_state_2011,milewicz_characterizing_2019}. 
It is now increasingly important for stakeholders involved in the sciences to ensure and improve the \emph{sustainability of scientific software}, defined as ``the ability to maintain the software in a state where scientists can understand, replicate, and extend prior reported results that depend on that software''~\cite{trainer2014community,lamprecht2020towards}. 
In the past decades, substantial financial resources have been allocated to developing and sustaining scientific software in different countries~\cite{katz_2021,alliance_rs_report,alliance_web_RS,canarie_RS,UK_SSI,horizon_EU,EVERSE,horizon_RS_EVERSE}, 
highlighting the significance of scientific software and the ongoing focus on sustainability concerns.

One primary challenge highlighted in previous research is that scientists generally lack sufficient training in software engineering (SE), which compromises the quality, usability, and sustainability of scientific software~\cite{merali2010computational,lowQualitySciSW2014,stodden2013best}. 
This challenge becomes increasingly important as the software evolves to be more complex. 
To address such a challenge, experienced software engineers are frequently brought on to the team to handle crucial engineering tasks~\cite{segal2009software,segal_scientists_2008,morris2009some,carver_software_2007}.
However, \emph{interdisciplinary collaboration} between scientists and software engineers can be difficult because the two groups often have different objectives (\emph{research} vs. \emph{development}), educational backgrounds, and mindsets~\cite{storer_bridging_2018,segal_scientists_2008,paine_who_2017,heroux_overview_2005}, thereby hindering 
the software development process~\cite{kelly_scientific_2015,hine_databases_2006,ribes_planning_nodate,lawrence_walking_2006}. 
Hence, the pursuit of establishing \textbf{\emph{sustainable collaboration}} in the development of scientific software remains a continuous area of research.
Meanwhile, in response to the Open Science Announcements
~\cite{federal_open} and other similar calls, more and more scientific communities have adopted open collaboration in their software projects~\cite{russell_large-scale_2018,schindler_role_2022,bangerth_what_2013,ahalt_water_2014,cos}. 
Given that the \emph{open-source model} has been successful in software development, there has been a steady growth in the use of scientific open-source software (OSS) and traditional scientific software teams are now transforming into large, diverse open-source communities.
Therefore, the challenge of sustaining scientific software lies in the need to cultivate \emph{\textbf{sustainable communities}} that take responsibility for its ongoing development and maintenance. 

It is important to also note that software projects do not exist in isolation; instead, they frequently constitute an ecosystem, which consists of a ``core that provides basic functionality'' and the surrounding user-contributed software packages~\cite{german2013evolution,bosch2009software,bosch2010integration,jansen2009sense,messerschmitt2003software}.
Accordingly, it is not unexpected that various scientific OSS also shape ecosystems that enable large-scale scientific discoveries. 
Notable examples include the \emph{ImageJ ecosystem}~\cite{imageJ_web}, 
the \emph{Bioconductor project}~\cite{Bioconductor_web}, and
 the generation of the first image of a black hole facilitated by more than 11 software packages from the scientific Python ecosystem\cite{python_sic_web, blackhole_case}.
As various interoperable projects come together to form a larger scientific OSS ecosystem, effective collaboration among these projects becomes essential. Tackling \emph{\textbf{the challenge of sustainability at the ecosystem level}} requires non-trivial cross-project communication and coordination.

Previous work has investigated the challenges of interdisciplinary collaboration between scientists and software engineers 
in developing in-lab or commercial software ~\cite{kelly_scientific_2015,paine_who_2017,segal_scientists_2008} and proposed strategies to improve collaboration efficiency. Examples of these strategies include increased education~\cite{widder_barriers_2019}, effective interaction~\cite{arvanitou_software_2021}, and hosting hackathons to facilitate knowledge exchange~\cite{taylor_understanding_2019}. 
Yet, it is unclear if the challenges identified and solutions proposed still hold in open-source settings where collaboration primarily takes place asynchronously, remotely, and among a vast and diverse group. 
Likewise, previous research on motivations for contributing to OSS has examined reasons for contributor disengagement and proposed solutions to enhance the sustainability of OSS communities, but it applies more generally to traditional OSS. 
The fundamental differences between scientific software and OSS in terms of their funding model and stakeholder composition~\cite{robinson_sustainability_nodate} raises doubts as to 
whether scientific OSS shares similar concerns about community sustenance \cite{raman2020stress,singh2021codes}
as traditional OSS. 
Furthermore, it is questionable
 whether the existing solutions in OSS (e.g., effective governance of OSS community, sponsors and donations to support developers~\cite{gamalielsson2014sustainability,shimada2022github}) 
can be directly applied to scientific OSS communities. 
Note also that while several authors have studied the sustainability of multi-project software ecosystems
~\cite{burstrom2022software,manikas2016revisiting,manikas2013software}  
and OSS ecosystems~\cite{franco2017open} by covering a range of subjects (e.g., the processes, challenges, health measures~\cite{wang2019toward}, and evolution~\cite{german2013evolution}), 
they primarily focused on the code dependency relationship~\cite{canfora2011social,ma_how_2017, chen2022collaboration,tan2022exploratory} 
without investigating how different projects collaborate. 
Including the scientific domain of the OSS ecosystem in the discussion of sustainability will help determine if the findings for traditional OSS are consistent with scientific OSS ecosystems. 
Therefore, we argue that the convergence of three characteristics of scientific OSS, namely \emph{interdisciplinary collaboration}, \emph{the open-source mechanism}, and \emph{multi-project collaboration},
prompts us to examine \emph{\textbf{whether the sustainability challenges identified within the context of each characteristic remain the same overall for scientific OSS}}.

Due to the wide range of scientific disciplines, it is challenging, if not impossible, to identify a group of 
software projects that represent the entire landscape of scientific OSS and their corresponding communities. 
Thus, rather than targeting generalizability to software development in all scientific domains,
our intention is to first perform a case study focused on a specific scientific OSS ecosystem that investigates the three characteristics mentioned earlier. 
Considering the exploratory nature of our research objective, undertaking a case study on the development process of a scientific OSS would offer a real-world example to gain a thorough understanding of the challenges~\cite{stol2018abc}.
In this study, we chose the \projectname{} project~\cite{astropy}, 
a software ecosystem comprising one core package and 50 interoperable packages (organized as separate repositories on GitHub and maintained by different teams as of August 2022) that offer commonly used functionalities in Python for astronomy. The core package (\url{github.com/astropy/astropy}) has over 1.7K forks, showcasing its widespread popularity (detailed justification is provided in Sec.~\ref{astropy-intro}). 
Correspondingly, we ask the following three research questions (RQs):
\begin{itemize}
      \item  \textbf{RQ1 (interdiscipline)}: \RQone{} 
      We study the collaboration challenges faced by scientists and software engineers in an open-source setting, which would impede the long-term sustainability of both the software and the community.
    \item  \textbf{RQ2 (open-source)}: \RQtwo{} Given the notable distinctions between scientific OSS and traditional OSS (details are described in Sec.\ref{related_scioss}), 
    we would like to understand if there are unique factors concerning the sustainability of the OSS community in the scientific domain. 
     \item  \textbf{RQ3 (ecosystem)}: \RQthree{} 
     We characterize the current cross-project collaboration practices and challenges, considering them as crucial elements influencing the sustainability of the ecosystem as a whole.
  \end{itemize}

To answer the RQs, 
we adopted a \emph{mixed-methods design} approach~\cite{flyvbjerg2006five,guetterman_two_2018}: 
We first interviewed core developers of \projectname{} project with different backgrounds to understand the collaboration practices and challenges in interdisciplinary teams. 
Then, we surveyed disengaged contributors to understand their motivations for contributing and reasons for disengagement from the \projectname{} communities. 
Meanwhile, we analyzed all 51 repositories in the \projectname{} ecosystem to triangulate our findings from interviews and surveys. 
Finally, we leverage network analysis techniques to study the cross-referenced issues and pull requests (PRs) across various projects to discern the rationales behind cross-project collaboration and compile potential opportunities for enhancing collaboration efficiency within the ecosystem. 

Our results indicate that while the \projectname{} project as a scientific OSS encounters many similar development obstacles to traditional OSS, 
the high entry barrier of the scientific field and interdisciplinary team composition seriously challenge the sustainability of both the software development process and the community. 
At the ecosystem level, apart from the frequently observed code dependency relationships among projects and the necessity for coordination to prevent breaking changes, 
we observe other forms of collaboration. 
These encompass coordinating centralized infrastructure management, common functionality, and knowledge sharing. 
Drawing from our findings, we propose best practices and future tooling and research directions to foster efficient collaboration and sustained development for the \projectname{} project and scientific OSS in general.

\section{Related Work}\label{related}

\subsection{Scientific OSS}\label{related_scioss}

\textit{\textbf{Increasing awareness of the importance of scientific OSS.}} 
The growing importance of scientific OSS in recent years 
is reflected by the increasing number of prominent funding agencies launching dedicated programs to support the maintenance, common infrastructure, and community engagement of scientific OSS. 
For example, the Chan Zuckerberg Initiatives (CZI) established the ``Essential Open Source Software for Science'' program~\cite{zuckerberg_esss}, and the Sloan Foundations funded the ``Better Software for Science'' program~\cite{sloan_bss}. 
Similarly, long-term commitment initiatives are created to support scientific OSS from institutions, such as NASA's Open Source Science Initiative~\cite{nasa_opensci}. 
Organizations such as Better Scientific Software (BSSw)~\cite{bssw} and Research Software Alliance (ReSA)~\cite{resa} have been established to promote best practices in scientific software development and provide training and resources for scientists to improve their software development skills; 
NumFOCUS~\cite{numfocus_web} has been dedicated to locating financial support for open-source scientific software projects. 
Practices like citation support on coding platforms such as GitHub~\cite{github_citation} and publishing papers about research software (e.g., Journal of Open Source Software (JOSS)~\cite{joss_web}) has been established to acknowledge and give credit to the contribution of scientific software in current academic systems. 
However, they also highlight the remaining concerns and the urgent need for better support and practices to develop high-quality and sustainable scientific OSS.

\textit{\textbf{Scientific OSS communities publish papers to share their experiences in managing communities, as well as challenges and practices related to sustainability.}} 
This is a great way to promote the software and encourage users to cite the software to credit the contributors in the academic rewards system. 
For example,
the NumPy community published a paper describing the 
designs, the example use cases in scientific research,
its contributors and funding composition, 
as well as the role of the software in the broader ecosystem around it~\cite{harris2020array}; 
the SciPy community has a paper addressing similar themes~\cite{virtanen2020scipy}; 
the Astropy community published three papers over~\cite{the_astropy_collaboration_astropy_2013,price2022astropy,price2018astropy} the nine years following each major update of the software, with the latest publication discussing the feature designs of the software, the roadmap for future development, the application of the software in the research domain, along the communities efforts on sustaining the software, including the community management, education plans, the governance structure, and the funding plans~\cite{price2022astropy}; 
rOpensci~\cite{ropensci_web} community's publication shared their experience on addressing both technical and social challenges in building sustainable software and community, 
including active social media engagement, education for domain scientists users in the forms of workshops, hackathons, and other social events, 
as well as recommending standardized domain-specific workflow for scientists to address the reproducibility concerns~\cite{boettiger2015building}. 
To the best of our knowledge, the 2019 technical report by CS\&S (Code for Science \& Society) is  the closest to our study goals~\cite{robinson_sustainability_nodate}, 
which summarized surveys
 and interviews with contributors and users 
over eight research-driven OSS communities.
 The study identified three key themes regarding the sustainability challenges: funding, leadership, and maintenance. 
However, the report only studied the existing community members and mainly focused on management. 
Our research complements the mentioned paper by conducting an in-depth case study, 
investigating sustainability challenges from a collaboration perspective and within the surrounding scientific oss ecosystem.

\textit{\textbf{Prior studies on scientific OSS within the software engineering communities.}} 
Apart from the papers submitted by scientific OSS communities themselves, the SE research communities have begun to examine the software from various perspectives. 
For instance, 
Milewicz et al. undertook a study on seven scientific OSS projects to explore collaboration practices among various roles based on contributors' academic seniority, categorizing them as senior research staff, juniors, and third-party contributors~\cite{milewicz_characterizing_2019}. 
Codabux et al. investigated the technical debt types in the documentation of the scientific OSS packages in R programming language through a case study with the \emph{rOpenSci} ecosystem~\cite{codabux2021technical}.
Additionally, Sharma et al. developed a model to automatically detect different types of technical debts in the development process of R packages and empirically identified the causes of the technical debts~\cite{sharma2022self}.
In contrast to these studies, our focus is on sustainability challenges. 
We delve into three aspects -- interdisciplinary collaboration, the open-source environment, and the multi-project ecosystem -- to achieve a thorough comprehension of scientific OSS.

\subsection{Scientific Software Development}

\textit{\textbf{Sustainability challenges.}}
Earlier research has explored the technical hurdles related to developing complex scientific software~\cite{carver2016software} or specific challenges during development phases like design~\cite{queiroz2022science}, testing~\cite{kanewala_testing_2014}, and release~\cite{lin_releasing_2019}. 
They observe that scientific software often falls short of desired quality attributes~\cite{smith_raising_2021,taylor_understanding_2019,merali2010computational,lowQualitySciSW2014,pitt2008chaste,stodden2013best} such as traceability~\cite{hata2021science,wattanakriengkrai2022github}, reproducibility~\cite{widder_barriers_2019,krafczyk2019scientific,merali2010computational},
 and sustainability~\cite{trainer2014community,lamprecht2020towards}.

Given scientific software's vital role in research success, 
funding agents and research institutions emphasize the necessity for sustainable software over an extended period~\cite{downs2015community,katz2015report,bssw,cssi,sloan_bss}. 
Numerous suggestions aim to enhance programming practices for scientific software, including providing training to scientists to improve their programming skills~\cite{stodden2013best,merali2010computational,widder_barriers_2019,milewicz_characterizing_2019}, and building better software and maintenance infrastructures
~\cite{kelly_scientific_2015,storer_bridging_2018,smith_raising_2021,taylor_understanding_2019}.  
Recent survey and systematic mapping studies provide a comprehensive summary of previous work on applying SE practices in scientific programming 
~\cite{storer_bridging_2018,arvanitou_software_2021}. 
Despite repeated investigation and reporting of challenges, particularly related to collaboration and reproducibility, they persist and remain unresolved~\cite{ivie2018reproducibility,krafczyk2019scientific,pimentel2019large,koehler2020better}.

\textit{\textbf{Interdisciplinary collaboration.}}
During software development, team members with varying educational and professional backgrounds tend to have different expectations of the system, 
leading to longer communication and coordination efforts~\cite{brandstadter2016interdisciplinary,brooks1974mythical}. 
The interdisciplinary nature of scientific software often requires an even more collaborative environment~\cite{milewicz_characterizing_2019,arvanitou_software_2021,howison_scientific_2011}, including both collaborations between scientists from different fields~\cite{ahalt_water_2014,howison_scientific_2011,storer_bridging_2018} and between scientists and software engineers~\cite{kelly_scientific_2015}. 
Thus, developing scientific software requires a specialized skill set beyond the general knowledge and experience in software development~\cite{bangerth_what_2013}. 
Prior work has studied specific interdisciplinary collaboration challenges between software engineers and data scientists ~\cite{nahar2022collaboration}, 
and between software engineers and user experience (UX) designers~\cite{maudet2017design,almughram2017coordination}. 
Unlike in those work where each group of experts has distinct responsibilities (e.g., \emph{model training} vs. \emph{building the system}, \emph{design} vs. \emph{implementation}), the boundary between scientists and software engineers becomes less clear when developing scientific software. 
In fact, the majority of development work is undertaken by scientists first and as the system grows more complex and requires maintenance, professional software engineers are hired~\cite{bangerth_what_2013}. 
Moreover, existing work focuses on the collaborative development of scientific software, often led by research labs or institutes. 
However, interdisciplinary collaboration in an open-source environment, where software is developed mainly through voluntary community efforts, remains unexplored. 
Therefore, we aim to investigate the unique challenges of developing scientific software in such an environment and contribute valuable insights to collaborative SE.

\subsection{Sustaining OSS communities and ecosystems} \label{related-oss-sustain}
\textit{\textbf{Retaining contributors.}}
Because of the voluntary nature of contributions, OSS projects have been facing sustainability challenges regarding retaining existing contributors and attracting newcomers~\cite{chengalur2010sustainability}. 
Prior studies investigated the motivation of different types of contributors, including long-term core contributors and peripheral contributors who make casual contributions~\cite{krishnamurthy2016peripheral}. 
Various types of motivations have been observed, including (1) intrinsic motivations, such as altruism, kinship, fun, and ideology; (2) internalized extrinsic motivations such as reputations, 
reciprocity, learning, and own-use (e.g., ``scratch one's own itch'')~\cite{alexander2002working,lakhani2003hackers,ghosh2002free}; 
and (3) extrinsic motivations, such as career and pay~\cite{von2012carrots}.
In 2021, Gerosa et al. revisited the motivations for contributions previously identified in research on OSS projects to examine how motivation has evolved in light of the emergence of platforms like GitHub~\cite{gerosa2021shifting}.
They found a growing emphasis on social motivations over the years. 
Still, when contributing to OSS projects, newcomers face both social barriers (e.g., lack of social interaction with maintainers) and technical barriers (e.g., lack of necessary previous knowledge, difficulty finding tasks to work on)~\cite{bird2011sociotechnical}.
Moreover, the disengagement of existing contributors due to reasons like heavy workloads~\cite{miller2019people} or hostile community culture~\cite{gray2022disengage} poses another challenge to the sustainability of OSS projects.

\textit{\textbf{Lowering the entry barriers.}} Best practices such as good first issues~\cite{tan_first_2020}, mentoring~\cite{fagerholm_role_2014}, and the Google summer of code programs~\cite{silva_theory_2020} have been adopted by OSS projects to help onboard newcomers. 
At the same time, many projects have leveraged donations and funding from organizations to better support the maintenance process and retain existing contributors~\cite{shimada2022github,overney_how_2020}. 
Other practices such as the Code of Conduct are also widely adopted to facilitate welcoming and healthy community cultures~\cite{singh2021codes}.

As a subset of the OSS communities in general, scientific OSS face the commonly mentioned sustainability challenges, such as attracting newcomers~\cite{kraut2012building,dominic2020conversational}, retaining contributors~\cite{stanik2018simple}, scaling up the communities~\cite{tan2020scaling}, etc.
Nevertheless, it is unknown whether 
the concerns in sustaining communities are common to both scientific OSS and general OSS. 
Additionally, it is unclear whether  
the solutions proposed in previous work are still applicable, considering the distinct characteristics between scientific OSS and general OSS, such as:
\begin{itemize}
    \item Various motivations drive contributions. Academic credit is assigned to scientific contributions, while open-source contributions are gauged by the effort put into software-related activities~\cite{howison_incentives_2013}. 
    \item It is more challenging to onboard and retain contributors to scientific projects than traditional OSS as explored in academic research and grey literature studies~\cite{price2022astropy, Accelerate_Sci}. 
    \item Different funding models, such as research institute affiliations and grant funding, differ significantly from the predominantly voluntary approach commonly observed in general OSS projects~\cite{kelly_scientific_2015,paine_who_2017}
    \item Distinct stakeholder compositions, exemplified by a small community affiliated with a specific research domain versus a diverse contributor background, contribute to the differentiation between the two~\cite{kelly_scientific_2015,paine_who_2017}
\end{itemize}

\noindent
Note that part of our study (RQ2) can be considered as a \emph{conceptual replication}~\cite{juristo2012replication,schmidt2016shall} of prior work, which concentrates on the sustainability challenges of general OSS communities\cite{gerosa_shifting_2021}, focusing on the motivation of contribution 
and the reasons for disengagement.
Our objective is to identify the pain points in the context of scientific OSS that may not be adequately tackled by current solutions.

\textit{\textbf{Studies about OSS ecosystems and cross-project coordination.}}
Frequently, multiple OSS projects form larger ecosystems, such as the Eclipse ecosystem~\cite{eclipse_ecosys}, 
R ecosystem~\cite{CRAN_site,CRAN_packages}, 
scientific Python ecosystem~\cite{harris2020array}, 
and the \emph{NPM} ecosystem~\cite{wittern2016look}. 
Cross-project collaboration is characterized by complex contextual interactions that manage tasks, 
activities, and interdependencies across multiple projects~\cite{university_of_auckland_artifacts_2010}. 
Researchers have compiled the existing studies on OSS ecosystems via systematic mapping studies, highlighting various areas of investigation, such as the licensing model, the co-evolution, co-creation, and the market-places~\cite{franco2017open,manikas2013software,ameller2015development}.
When examining the scenario of cross-project coordination, the majority of the studies concentrated
on the code dependency relationship 
(e.g., cross-project bug fixing~\cite{canfora2011social,ma_how_2017, chen2022collaboration}, breaking changes~\cite{BogartKHT21}, software supply chain~\cite{tan2022exploratory,kikas2017structure}, and  the security risks~\cite{wu2023understanding}), 
and the practices of cross-project code reuse~\cite{gharehyazie2017some}. 
Additionally, various methods were designed to improve coordination efficiency such as 
impact analysis of cross-project bugs and detecting the affected downstream modules~\cite{ma2020impact}, and dependency management tools to improve the quality of the dependency network~\cite{hejderup2018software}, etc. 
Moreover, researchers also leveraged social network analysis to illustrate the relationship among developers and projects aiming to support knowledge collaboration across projects in the open-source settings~\cite{ohira2005accelerating,herbold2021systematic}. 
While our work also explores cross-project coordination, 
our focus is to understand its role in ensuring the sustainability of a scientific OSS ecosystem.

Prior work also explored domain-specific OSS ecosystems like ROS (an ecosystem of packages for robotics development~\cite{ros_web}). 
For example, Alami et al. investigated factors influencing the quality assurance (QA) practices in the ecosystem~\cite{alami2018influencers}. 
They observed that the absence of a working sustainability strategy and the interdisciplinary nature of the system pose obstacles to the successful implementation of QA. 
Similarly, Kolak et al. studied the ROS ecosystem, exploring the influence of domain-specific knowledge on the open-source collaboration model, the structure of the collaboration, and the evolution of dependency networks over time ~\cite{kolak2020takes}. 
In contrast to these studies, our research delves deeply into sustainability challenges, examining the factors that limit sustainability, such as interdisciplinary collaboration, the open-source environment, and the ecosystem. 
Moreover, we study the connections between sub-communities beyond the dependency relationship by incorporating various forms of collaborative intentions.

\section{Research Design}

We conduct an exploratory case study to examine sustainability-related issues from three aspects -- interdisciplinary collaboration (RQ1), open-source mechanism (RQ2), and multi-project ecosystem (RQ3). 
In this section, we first provide a rationale for the chosen study subject, followed by an introduction to the research methods.

\subsection{Subject Selection: the Astropy Project} \label{astropy-intro}

\begin{figure}[t]
\centering
\footnotesize
\includegraphics[width=.7\textwidth]{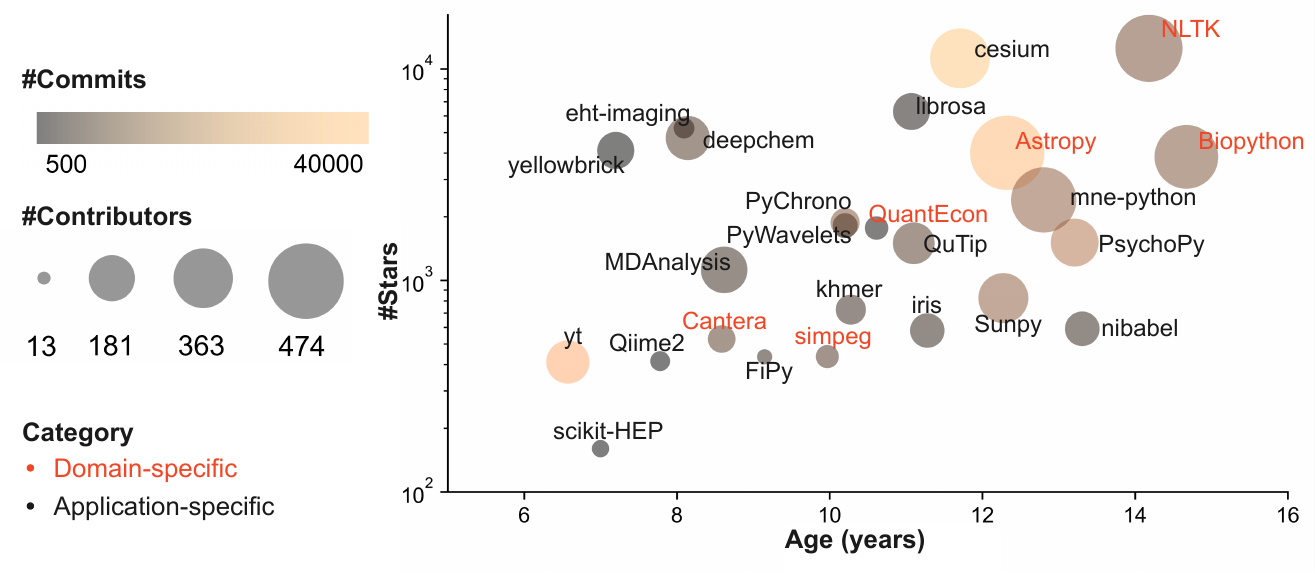}
\caption{The scientific Python ecosystem. 
For each dot, the size represents the number of contributors to the package, the color gradient indicates the commits count of the package, and the color of the text labels demonstrates the category of the packages.} 
\label{fig:numpy_tree}
\end{figure}

For the choice of the case subject, we leverage the existing structure of the scientific Python ecosystem depicted by previous research~\cite{harris2020array}, 
in which NumPy~\cite{numpy_web}, SciPy~\cite{scipy_web}, and Matplotlib~\cite{matplotlib_web} serve as the foundation of the whole ecosystem, and the rest of scientific OSS is categorized into three layers: 
(1)  \emph{Technique-specific}: scikit-learn, scikit-image, pandas, statsmodels, NetworkX;
(2)  \emph{Domain-specific}: Astropy, Biopython, NLTK, QuantEcon, cantera, simpeg;
and (3)  \emph{Application-specific}: cesium, FiPy, yellowbrick, MDAnalysis, etc. 
We aim to select a project within the domain-specific category that fulfills the three specified aspects. 
We consider the domain, maturity, popularity, available development artifacts as well as the surrounding local ecosystem of each software as the criteria for the selection. 
We compare these projects in both \emph{domain-specific} and \emph{application-specific} categories based on four dimensions: project age, commit count, contributor count, and star count, as shown in Fig.~\ref{fig:numpy_tree}. 
Eventually, we choose the \projectname{} project~\cite{astropyproject}, an ecosystem of software packages\footnote{To avoid confusion, we will use ``package'' throughout the paper, as it is often used interchangeably with ``project.''} for astronomy, as the subject because it is a popular and mature project with the largest number of contributors and commits. 
With over 18,461 citations of the papers published by the community~\cite{the_astropy_collaboration_astropy_2013,price2018astropy,price2022astropy} over 10 years, \projectname{} project is widely used by researchers and practitioners in astrophysics research as well as scientific missions, such as James Webb Space Telescope~\cite{jwst_web} and Event Horizon Telescope Collaboration~\cite{EHT_Web,blackhole_case}. 
The code repositories and development artifacts of the packages in the \projectname{} project ecosystem are publicly available via GitHub and official documentation.
This allows us to study tangible issues in a rich context related to sustaining software projects, fostering communities, and enhancing collaboration efficiency throughout the ecosystem.

\textbf{The community.}
The community members could be divided into: (1) the \emph{official Astropy Team}~\cite{astropy-teams}, comprising individuals with administrative roles and responsibilities, involving tasks related to code contributions (e.g., repository maintenance, development \&operations specialist) or non-code contributions (e.g., handling finances, community engagement); and (2) contributors, who do not hold official roles but actively contribute to the repositories through various means (e.g., code contributions, reporting issues).

\textbf{The software ecosystem.} The \projectname{} project is an ecosystem including 51 computing-related Python packages (as of August 2022) and multiple infrastructure packages~\cite{price2022astropy}.
For the computing-related packages, they could be further categorized into three groups: 

\begin{itemize}
    \item The \emph{core} package, \emph{astropy/astropy}, provides 
    common functionalities such as data transformations, basic computations, and logging system~\cite{astropygithub}. It is maintained by ``core-package maintainers''~\cite{astropycoremaintainer}, who are part of the official \emph{Astropy Team}. 
    \item 7 \emph{coordinated packages}, that are either ``experimental or 
   problem space-focused'' not fitting into the core package at that moment~\cite{astropycoordinatedpkg}. 
    They are developed and maintained by the 
    ``coordinated package maintainer'', also part of the offical \emph{Astropy Team}.
    \item 43 \emph{affiliated packages}, are astronomy-related packages offering functionalities that are not part of the core package. These packages are not under the management of the official \emph{Astropy Team}; rather, they are maintained by distinct teams, forming diverse sub-communities within the ecosystem~\cite{astropyaffiliatedpkg}.
\end{itemize}

\noindent
Note that the code base of the packages is continually evolving. 
For example, if a functionality within a \emph{coordinated package} becomes widely used across numerous cases, the code may be incorporated into the \emph{core package}. Likewise, if an \emph{affiliated package} proves valuable to a broader community, the \emph{official Astropy Team} may take over the maintenance responsibility and elevate it to a \emph{coordinated package}. 
Over the past 10 years, the \projectname{} project has transformed from a single core package \emph{astropy/astropy} and progressively grown into a comprehensive ecosystem, consisting of various interoperable packages that either build upon or extend the functionalities of the core package~\cite{price2022astropy}. 
With the goal of promoting code reuse and collaboration, \emph{affiliated packages} joined the ecosystem after being reviewed by the \emph{official Astropy Team} members and meeting the requirements of the ecosystem guideline.

Apart from the 51 computing-related packages, there are several \emph{infrastructure packages}, maintained by the \emph{official Astropy Team},  that assist sub-communities within the ecosystem in easily building and maintaining their packages, lowering the barrier for scientists with little SE experience~\cite{price2022astropy}.   
These infrastructure packages consist of pre-configured templates, build tools, documentation generation, and testing toolkits~\cite{astropy_official_infra}.

\subsection{Research Methods}

\begin{figure}[t]
\centering
\includegraphics[width=0.65\textwidth]{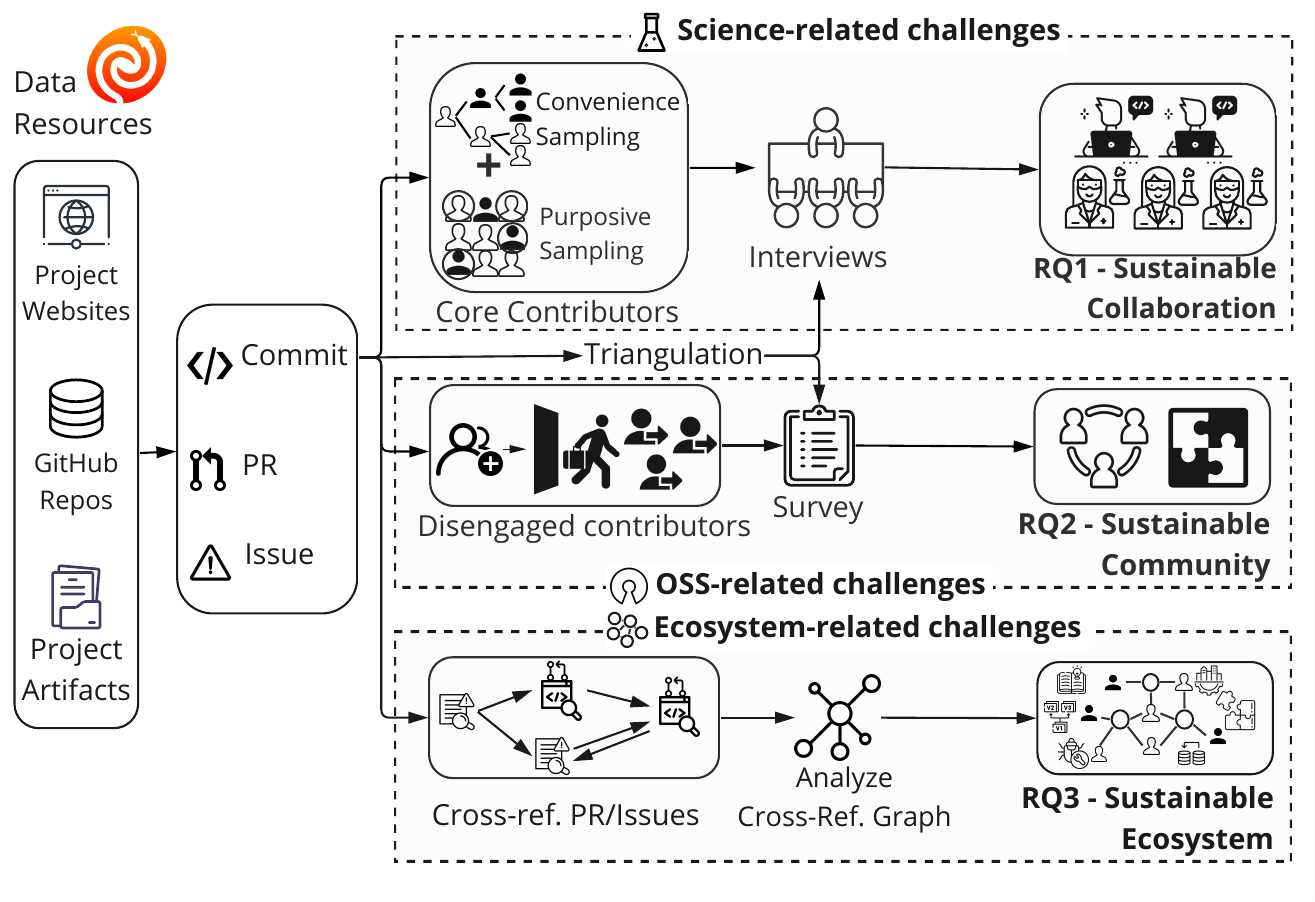}
\vspace{-1em}
\caption{Research Method Overview.}
\label{fig:method}
\vspace{-1em}
\end{figure}

We adopt a mixed-methods design~\cite{runeson_guidelines_2009,flyvbjerg2006five} including  mining software repositories (MSR), interviews, 
and surveys to both qualitatively and quantitatively understand the sustainability of \projectname{} from different aspects. 
The three types of methods are not used in a linear manner. 
Instead, we adopt an iterative approach to refine the study design, where the data collection, analysis, and results of these methods mutually inform one another~\cite{wohlin2012experimentation,runeson_guidelines_2009,yin2009case}. 
For example, \emph{MSR} serves many purposes during the study, including identifying the study participants for both interviews and surveys (RQ1\&2),  detecting cross-project collaboration (RQ3), and triangulating and complementing the qualitative findings (RQ1-3). 
Meanwhile, we gathered documentation from the official website to gain deeper insights into the software, the community, and the ecosystem.
The overview of our research method is presented in Fig.~\ref{fig:method}. 
 Specifically, the interview study is for qualitatively characterizing the challenges in collaboration and sustainability from long-term contributors' perspectives (RQ1). 
 The survey study aims to understand the motivations for contributing to scientific OSS as well as the reasons for disengagement (RQ2). 
The interview\&survey design material is available in the replication package~\cite{SupMt}, and was approved by the research ethics board of the authors' institution.

\subsubsection{\textbf{Interviews with Core Contributors.}}\label{interview}
Given the pivotal position of the \emph{core} package, \emph{astropy/astropy},  within the \projectname{} ecosystem, around which most other packages revolve, we consider core contributors of \emph{astropy/astropy} 
as key informants for our study as they have long-time involvement in the \projectname{} project, play a key role in its communities' evolution, and would have an in-depth view of these topics.

\textit{\textbf{Recruitment.}}
We identify 41 core contributors from the 424 contributors, considering their substantial contributions to the source code (responsible for 90\% of the total commits). 
Subsequently, we sent interview invitations to them using the email addresses collected from their public profiles. 
Meanwhile, we applied \emph{convenience sampling}~\cite{miles1994qualitative,ritchie2013selecting,shull2007guide} by requesting the interviewees to recommend other core contributors with diverse backgrounds who had not responded to our inquiries. 
Furthermore, we aim to include contributors from diverse backgrounds, such as researchers and professional software engineers, to gather comprehensive insights on various aspects (also known as maximum variation sampling~\cite{suri2011purposeful}).
In total, we received 12 responses and conducted 11 semi-structured interviews, and one was canceled due to scheduling issues. All interviewees 
have been involved in both software development and management activities of \projectname{}.

\textit{\textbf{Interview Design and Analysis.}} The interviews were done remotely on Zoom except for one, which was done in person, all lasting 45 to 60 minutes. 
The interview script underwent iterative revisions, involving the removal of irrelevant questions, the addition of new ones, and improvements to existing ones. 
There is no compensation for participating in the interview study.
During the interview, we asked participants about their roles in the \projectname{} community, their educational background in both SE and the scientific domain, their experiences in contributing to OSS projects, their opinions on interdisciplinary collaboration in the \projectname{} project and sustenance of its community, etc. 
Each interview was transcribed and analyzed immediately after completion. 
We then conducted thematic analysis~\cite{cruzes2011recommended} for analyzing the transcript of the interviews.
Each interview was independently analyzed by one of the first two authors,
adopting an inductive open coding approach to identify the themes regarding collaboration and sustainability challenges, as well as practices~\cite{cruzes2011recommended,saldana2021coding}. 
After the initial coding, the two authors discussed the coding results and reached a consensus on the themes.
To validate our perception of the interviewees' responses, we sent follow-up emails to our interviewees to clarify any ambiguous information, as well as the themes identified from the interviews~\cite{candela2019exploring,birt2016member}. 
We reached saturation after the fifth interview, which means no new themes are identified in the following interviews and the additional interviews provided only marginal new insights~\cite{francis2010adequate}.

\subsubsection{\textbf{Survey with Disengaged Contributors.}}\label{survey-sec} 
To gain supplementary information regarding the sustainability of the \projectname{} community (RQ2), 
we conducted a survey with contributors that disengaged from the \projectname{} ecosystem. 
The survey consists of open-ended questions~\cite{stol2018abc} to qualitatively identify their motivations for contributing to scientific OSS and the reasons for disengagement.

\textit{\textbf{Recruitment.}} Disengaged
contributors are defined as those who meet all of the following criteria:
\begin{itemize}
    \item The developer has merged at least one commit to any of the 51 packages in the past, and
    \item The developer has not made any contributions to the code base within the last 100 days or more, and
    \item The developer is still active in other OSS projects outside of the \projectname{} project ecosystem. 
\end{itemize}
This activity pattern serves as an indicator of having recently ceased or departed from participation in the \projectname{} community.  
Note that related work has explored different thresholds (e.g., 30, 90, 180 days)~\cite{milewicz_characterizing_2019,lin2017developer} and showed similar results.
The detailed procedure can be divided into three steps:

\begin{itemize}
    \item \emph{Step 1}: We analyzed the commit history of all 51 packages to identify commit authors, resulting in 1,116 GitHub accounts. 
    \item \emph{Step 2}: For each account, we collected their latest development activity (i.e., commits, PRs, and issues) via GitHub API and identified the potential disengaged contributors using the criteria described above. We further removed bot accounts and deleted accounts. This results in 469 GitHub accounts.  
    \item \emph{Step 3}: We further filtered out the ones without public contact information, resulting in 292 potential participants to whom we sent our survey invitation.
\end{itemize}

\noindent
\emph{\textbf{Survey Design and Analysis.}} 
 Three open-ended questions were asked in our survey: (1) motivations for contributing to \projectname{}, (2) reason for disengaging from the community, and (3) perception and suggestions for retaining the contributors and sustaining the \projectname{} community and the scientific OSS community in general. 
In total, we received \numsurveyee{} responses (response rate 23\%). 
To identify the salient themes related to community sustainability challenges from the perspective of disengaged contributors, we applied open coding and axial coding the analyze the textual responses~\cite{saldana2021coding}.

\subsubsection{\textbf{MSR for Constructing Cross References Graphs (CRGs).}} \label{network}
To gain a comprehensive understanding of the sustainability challenges from the ecosystem perspective (RQ3), with a particular emphasis on the dynamics of cross-project collaboration, we investigate the interaction among different communities through \emph{cross-reference links} during code-review and issue-resolving processes as instances. 
Note that both Li et al.~\cite{li_how_2018} and Hirao et al.~\cite{hirao2019review} both studied the linked issue/PRs on GitHub and OPENSTACK correspondingly to understand the types of the linking and the impact of linkage behavior on the code review process.
In this context, our research complements these two studies by examining the links from a collaborative perspective to understand the intention and derive future research and tooling suggestions.

\begin{figure}[t]
\footnotesize

\subfloat[Example: Cross-reference events connecting three projects.]{%
  \includegraphics[clip,width=0.6\textwidth]{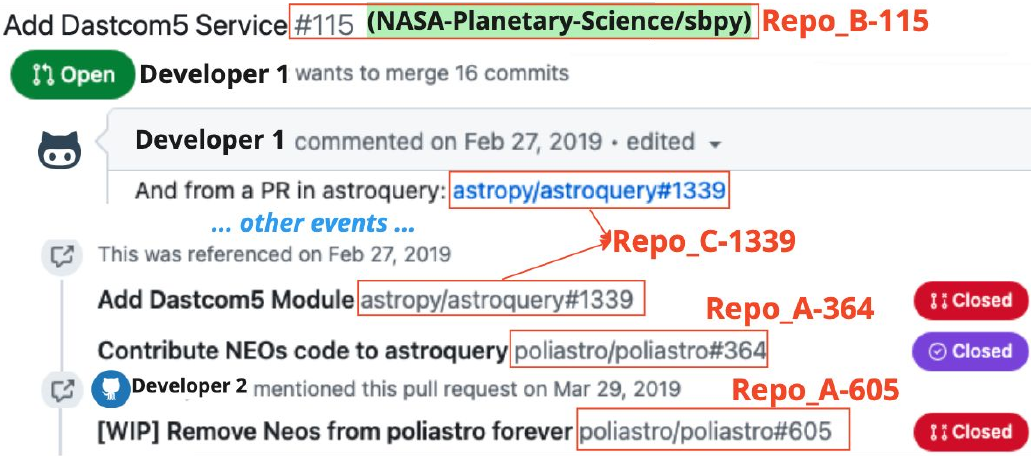}%
}
 \vspace{-1em}
\subfloat[Cross-Reference Graph (CRG) for the example above.]{%
  \includegraphics[clip,width=0.5\columnwidth]{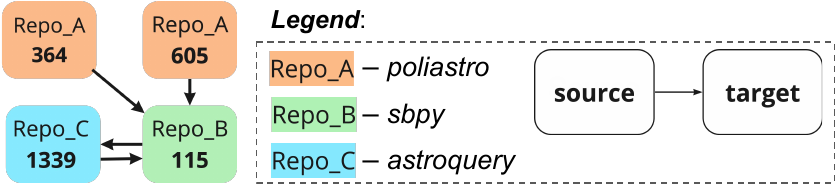}%
}
\footnotesize
\caption{Cross-reference events example and the corresponding Cross-Reference Graph (CRG), where the nodes represent issues/PRs, and edges depict connections between source and target nodes from three projects.}
 \label{fig:cross-ref-graph}
\end{figure}

\textit{\textbf{Collecting cross-reference links.}} 
GitHub offers cross-referenced links either within or across projects as directional links between a PR/Issue (source) to another PR/Issue (target) created during discussion. 
For example, Fig.~\ref{fig:cross-ref-graph}(a), 
presents the partial timeline events of \emph{PR\#115} in \emph{NASA-Planetary-Science/sbpy}, 
which has been cross-referenced three times from other two projects (\emph{poliastro/poliastro} and \emph{astropy/astropyquery}).
We collected all the cross-referenced links in the \projectname{} ecosystem by parsing all the issues and PRs using the GitHub Timeline events API~\cite{githubtimeline}, resulting in a total of 31,576 cross-reference links. \szmodifyok{across packages within the ecosystem}{} 
We then excluded links within the same project and retained only those connecting source and target nodes from different projects. If there are multiple links that exist from source to target, we only count one occurrence to avoid duplicated analysis. This results in a final count of 1,491 links across 43 distinct packages.

\textit{\textbf{Graph construction.}} 
Next, we construct a \emph{Cross-Reference Graph (CRG)}, 
where the nodes represent issues/PRs, and edges are cross-reference links
between source and target nodes.  
Note that the graph is not fully connected but includes several disjointed components (also known as sub-graphs) meaning some nodes cannot be reached from other nodes. 
In total, the 1,491 links contribute to the formation of \numCRGSubgraph~ CRG sub-graphs. 
In the rest of this paper, we call each ``CRG sub-graphs'' as CRGs for short. 
For example, Fig.~\ref{fig:cross-ref-graph}(b) illustrates part of a CRG as an exemplar, it demonstrates how four nodes and four edges are connected in a CRG to reflect the example in Fig.~\ref{fig:cross-ref-graph}(a).

\textit{\textbf{Analysis.}} 
With the CRGs and the corresponding discussions in the linked issues/PRs, we conducted a thematic analysis~\cite{cruzes2011recommended} to understand the intentions of cross-project collaboration as well as the challenges and practices.
For each CRG, the analysis process consists of the following steps:
(1) First, we read through the title, body, and all activities along the timeline of the linked issue/PRs in the CRG to understand the contextual information. 
(2) Subsequently, we systematically explored all the links in the CRG in a breadth-first manner to gain a comprehensive understanding of the reasons behind the connections among the nodes.
(3) We focus on the underlying problem-solving objectives and the overarching topic around the discussions, ultimately yielding a comprehensive understanding of the collaboration effort. 
(4) Further, we document the inferred ``collaboration intention'' as labels for the cross-reference pairs.
It is common to identify multiple labels in one CRG.
To reduce the possible bias of the analysis~\cite{runeson_guidelines_2009}, two authors first independently analyzed the 15 links in the CRGs with the steps described above and reached a consensus on the interpretation of ``collaboration intention'' identified. 
Then one author analyzed the remaining CRGs.

\subsection{Threats to Validity and Reliability}
In this section, we discuss threats to validity in our study design and interpretation of results.
We employed open coding to examine the responses from the interviewees and survey participants, and the discussion content mined in the \projectname{} ecosystem, which may have resulted in unintended bias. 
We minimized this potential bias by having multiple coders independently analyze the data and reach a final agreement. We also triangulated our qualitative analysis results with the version control data and code review history obtained from GitHub. 
Taking the procedure of analysis cross-reference links as an example, to minimize bias during the labeling process, two authors independently labeled 20 randomly selected links using the described steps. Subsequently, the team discussed any instances of disagreement and derived a codebook, with each code representing a specific type of collaboration intention. After reaching a consensus on the codebook definition, one author proceeded to label the remaining links (edges) and their corresponding CRGs. Whenever uncertainty arose, the team discussed the case until an agreement was reached. 

The risk of participation bias was introduced as the participants who took part in our studies did so voluntarily. 
Moreover, we used a convenience sampling method to identify interviewees to interview; this is at the risk of sampling bias. 
Although we mitigate this risk by interviewing contributors with different roles and responsibilities, there may be other factors shared by the interviewees that do not apply to the entire community.
Caution should be exercised when making generalizations beyond the participants who were interviewed and surveyed. It is also worth noting that our results may not apply to other ecosystems outside of \projectname{}.

\section{Result: Interdisciplinary Collaboration (RQ1)}
In this section, we first present the composition of the core contributors in terms of expertise in the scientific domain and software engineering. 
Next, we triangulate the interview findings through the results of the contribution analysis of the commit history. 
Finally, we delve into the concrete obstacles encountered  by our interviewees during their interdisciplinary collaboration.

\subsection{Interdisciplinary Team Composition.}

\subsubsection{\textbf{Contributors' backgrounds.}} 
All 11 interviewees are in the core decision-making team of \projectname{}, responsible for community management, maintaining the packages, making long-term plans for the Project, etc. They have extensive backgrounds in astronomy, with six of them holding a doctoral degree, one having completed a master's, and two being Ph.D. students in the Astrophysics domain. 
They are from seven different organizations, such as research institutes and/or universities.\footnote{To prevent the risk of inadvertently revealing the identity of our interviewees, we refrain from providing specific details of their demographic background. The \projectname{} coordination committee has provided consent for disclosing the project name and has reviewed the manuscript. Additionally, all interviewees are aware of the potential risk of being identified as being part of our study.}
Two interviewees were self-identified as software engineers and also acknowledged by others as the only few \emph{\textbf{professional software engineers}} among the decision-making team. 
They are mainly involved in DevOps, project maintenance, code review, resolving issues, etc. 
One is paid by an institute to maintain the \projectname{} project, and the other is contributing voluntarily. 

The rest of the nine interviewees have daily jobs as researchers and hesitated to label their \szmodifyok{contributions as any particular type}{roles in the \projectname{} community as either software engineer or researcher}. 
As one interviewee pointed out, it is more appropriate to ``\emph{think of it as a spectrum from pure researcher to a pure software developer. 
There are people at all parts of that spectrum.}'' 
Meanwhile, they are also \textbf{\emph{users}} of some \projectname{} packages during their daily research activities. 
Five of them are \textbf{\emph{co-founders}} of the \projectname{} project with the motivation to ``\emph{collaborate on having one standard for doing all these things}'',  as they found that the development of scientific software packages in their domain was fractured, with individuals developing their own tools to attain similar goals. 
Despite lacking formal training in professional SE, they have made substantial code contributions to \projectname{}. 

Note that there is a discrepancy between the paper published by the \projectname{} team in 2022~\cite{price2022astropy} and our results regarding the distribution of software engineers and scientists in the team. 
The \projectname{} paper focuses on the ``top contributors'' ranked by the number of commits which is easier to measure quantitatively, while our interviewees mainly refer to the \emph{decision-making team} in practice, where only a few software engineers are involved.

\subsubsection{\textbf{Triangulating interview results.}}

\begin{figure}[t]
\centering
\footnotesize
\includegraphics[width=0.6\textwidth]{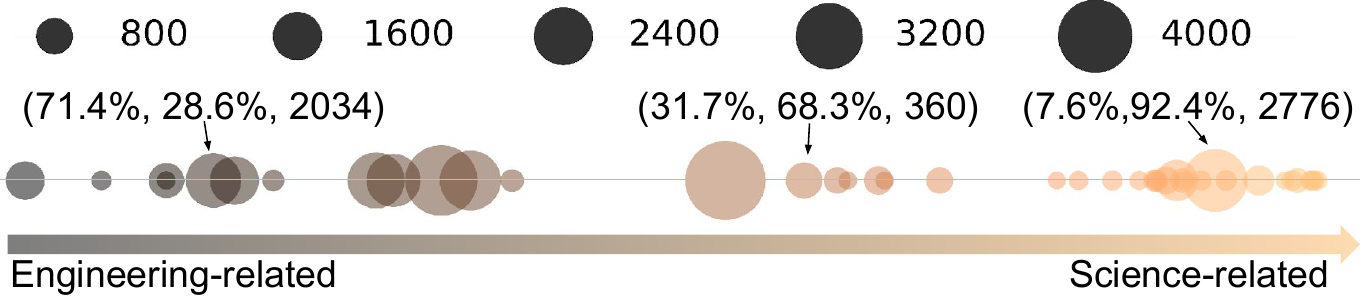}
\caption{Composition of contribution types of the 41 core contributors. Each dot represents a contributor and size reflects the number of commits. 
For each dot, $(i\%, d\%, c)$ represents the ratio of \emph{engineering-related} contribution, \emph{science-related contribution}, and \emph{the total number of commits}.} 
\label{fig:ratio_code}
\end{figure}

To further validate the \emph{spectrum of expertise} of the core contributors, we analyzed the contributions of 41 core contributors to the core package \emph{astropy/astropy} to approximate the contributors' backgrounds. 
According to the project documentation and also confirmed by our interviewees, the files in the code base can be grouped into two types: (1) \emph{science-related files} (e.g., source code of domain-specific functions, corresponding test cases and documentation), and (2) \emph{engineering-related files} (e.g., config file, utilities such as logging, or code of conduct), \szmodifyok{}{ which are perceived as more infrastructure\&management related}. 
For each contributor, we compute the {\emph{ratio of science-related contribution}} summing up the lines of code (LOC) contributed in science-related files over the entire commit history and then dividing it by their total number of contributed LOC. 
The result is shown in Fig.\ref{fig:ratio_code}, with the left side of the spectrum representing a higher contribution to \szmodifyok{software}{engineering}-related activities, and the right side indicating more science-related contributions.
The findings corroborated the interviewees' statements that contributors have a diverse range of expertise, and there is no strict demarcation between the roles of scientists and software engineers when it comes to the contribution preference on the core package \emph{astropy/astropy}.

\begin{tcolorbox}[boxsep=2pt,left=2pt,right=2pt,top=2pt,bottom=2pt]
\textbf{Finding 1}: 
A small proportion of the decision-making team of \projectname{} are professional software engineers -- some are paid to improve and maintain the repositories. 
All core contributors have a mix of scientific (professional) and SE backgrounds (mostly self-taught). Rather than exclusively focusing on one types of activities, their contributions often encompass a combination of both. 
\end{tcolorbox}

\subsection{Tension between Different Mindsets.}\label{tension}
 \everypar{\looseness=-1}
Given the ``spectrum of expertise'' in the core contributor team, it is not surprising that when asked about the communication challenges between scientists and software engineers,
our interviewees did not view the barriers between science and software engineering as an issue within the OSS context. 
As one interviewee suggested, ``\emph{there is always someone you can talk to who understands the domain knowledge of what you are working on but also has a lot of knowledge in the software structure}.'' 
Another interviewee also agreed that in open source most of the scientists who have engaged with the group are (at some level) thinking like engineers. 
In contrast, they encountered more notable challenges in their daily work environment outside of the OSS setting, where researchers and software engineers within the same organization 
frequently face communication breakdowns when conveying domain-specific functional requirements for the software. 

\szmodifyok{At the same time}{Nevertheless}, a noticeable discrepancy is observed between the \emph{mindsets} of the scientists and software engineers in the \projectname{} project. 
For instance, we observed a tension regarding \textbf{\emph{task prioritization}} between the two groups. 
On one hand, professional software engineers believe it is important to follow SE best practices and utilize automated workflows, such as continuous Integration and continuous Delivery (CI/CD), to reduce maintenance burden. 
As a result, they often explain to scientists that their code submission does not meet the code quality standard and needs refactoring. 
On the other hand, scientists perceive software engineers as ``\emph{who are not as familiar with the kind of flexible nature of scientific software collaboration}.''

When tension between mindsets arises, especially concerning coding standards, scientists in the team react differently. 
While some scientists are open to learning SE best practices, others demonstrate resistance. 
Those who refuse to adhere to the standards may become ``\emph{defensive and then just disappear and leave the pull request there to rot}'', as noted by one interviewee. 
This can substantially burden the software maintenance workload.
The resistance could be due to their primary focus on research rather than software development, resulting in limited time for code refactoring to meet contribution requirements. 
One interviewee pointed out that scientists tend to ``\emph{contribute as needed}'' and then ``\emph{go off to do other research until that happens again.}'' 
The software engineers in such cases have to make the hard choice of either taking over the code and improving it or ignoring the PR. 
Moreover, several interviewees pointed to long-standing issue discussions on GitHub as evidence of the tension between different mindsets. 
For example, in one discussion thread~\cite{astropyDiscussionBalance},
one maintainer expressed concerns about the current development infrastructure and its fragility, urging prioritization of maintenance work due to the increasing complexity of the ecosystem. 
Meanwhile, other contributors raised concerns over the complexity of workflow, which could steepen the learning curve for scientists and discourage their contributions.

Another tension revealed in our interview is the \textbf{\emph{perception of seniority}}. 
According to one interviewee who has experience in both \projectname{} and other OSS communities, one distinction between these two is the ranking of seniority -- in traditional OSS, contributors are ranked by the volume of their code contributions~\cite{vasilescu2015gender}, while in \projectname{}, people with a senior academic title could have more decision-making power on whether to merge PRs. 
Other interviewees also observed that ``\emph{senior researchers in the decision-making position sometime ignore certain PRs because they do not see the value of the research}.''
Furthermore, as described by some interviewees, sometimes the contributors who submitted PRs are not informed about the reason for keeping the PR open or rejected, which can be frustrating. 
One also mentioned that despite the management team’s “do-ocracy” policy~\cite{price2022astropy},
 which aims to empower those actively contributing to the project and ensure fair decision-making, there is still a noticeable difference between individuals with a scientific mindset and those with an engineering focus.
This unequal distribution of power can result in significant threats to the inclusiveness of the community, which we explore further in RQ2.

\begin{tcolorbox}[boxsep=2pt,left=2pt,right=2pt,top=2pt,bottom=2pt]
\textbf{Finding 2}: 
We observed two instances of tension between different mindsets: (1) \emph{Task prioritization} -- software engineers believe that SE best practices are often undervalued, while scientists desire more flexibility in balancing the importance of new scientific features and engineering tasks; and (2) \emph{Perception of seniority} -- unlike traditional OSS, where contributors are based on the quantity of their code contributions, the scientific OSS community incorporates an additional layer of academic seniority that influences the decision-making process.
\end{tcolorbox}

\section{Result: Sustaining Scientific OSS Community (RQ2)} 
In this section, we investigate the community sustainability factors and challenges in scientific OSS. 
As mentioned before, this part of the study could be considered as a \emph{conceptual replication}~\cite{juristo2012replication,schmidt2016shall} of prior work, but with a focus on scientific OSS communities.
Overall, we recognize challenges within the scientific OSS community across four aspects: \emph{onboarding newcomers}, 
\emph{retaining contributors}, 
\emph{improving rewarding system}, 
and \emph{resolving concerns on Equity, Diversity, and Inclusiveness (EDI)}. 
While some challenges are frequently observed in other OSS communities, others are specific to scientific OSS and may pose significant threats to the sustainability of both the software and its community.

\subsection{Onboarding \& Retention}\label{onboarding}\label{retention}
Related to onboarding newcomers and retaining existing contributors, we first summarize the motivations of people contributing to \projectname{} and the reasons of their disengagement based on our interviews and surveys. 

\begin{table}[t]
\begin{threeparttable}[b]
 \setlength{\tabcolsep}{0pt}
\caption{Motivation for contributing to \projectname{}. Note that the listed motivations are not mutually exclusive. One survey respondent can mention multiple motivations for contribution and would be counted for each category.}

\label{tab:incentives}
\footnotesize
\centering
\setlength\extrarowheight{0pt}
\begin{tabular}{lllr} 
\toprule
\textbf{Motivations }
& \textbf{~Description}
& \textbf{~Example response}
& \textbf{~\#\tnote{}}  \\ 
\hline
Own-use                           
& \begin{tabular}[c]{@{}l@{}}~ During research, edu-,~\\~cation, hobby usage\end{tabular}
& \begin{tabular}[c]{@{}l@{}}~~``\emph{...those integration functions were essentialfor my research,}\\\emph{ I developed  and committed } \emph{onto \projectname{}.}'' (SP63) \end{tabular}
& 47
\\
\hline
Altrusim
&~\begin{tabular}[c]{@{}l@{}}Benefit others via OSS\end{tabular} 
& \begin{tabular}[c]{@{}l@{}}~``\emph{Sharing my work so others can benefit} \emph{ thereby.}'' (SP61)\end{tabular}
&~11
\\
\hline
Learning
&~\begin{tabular}[c]{@{}l@{}}Gain experience via OSS\end{tabular} 
& \begin{tabular}[c]{@{}l@{}}~``\emph{Learn how to structure my programming} \emph{ better.}'' (SP66)\end{tabular}
&~11
\\
\hline
Invitation                           
& \begin{tabular}[c]{@{}l@{}}~Via OSS Meet-up,\\~or research workshop \end{tabular}
& ~\begin{tabular}[c]{@{}l@{}}~``\emph{It was part of a Hacktoberfest event and}\\\emph{I knew someone involved in the project.}'' (SP11)\end{tabular}
& ~4
\\
\hline
Pay                           
&~As SDE or researcher
& \begin{tabular}[c]{@{}l@{}}~``\emph{I was employed as a programmer in a lab.}'' (SP4)\end{tabular}
& ~5
\\
\hline
GSoC
& \begin{tabular}[c]{@{}l@{}}~Google Summer of Code \end{tabular}
& \begin{tabular}[c]{@{}l@{}}~``\emph{I was looking for a summer intern, so I}\\\emph{started contributing to \projectname{} for GSoC.}'' (SP55)\end{tabular}
& ~8
\\
\bottomrule
\end{tabular}
\end{threeparttable}
\end{table}

\textbf{Motivation for contributing to \projectname{} ecosystem.}
Our study aims to contrast the challenges of scientific OSS communities with traditional OSS. Therefore, after the open-coding process, we compared our results with the 10 motivations presented by~\citet{gerosa_shifting_2021}. 
Our results did not uncover any novel themes. Hence, we adopt the same terminology as \citet{gerosa_shifting_2021} used (see Tab.\ref{tab:incentives}). 
Given the space constraints, we only discuss the noteworthy findings or those that differ from the related work.

In contrast to Gerosa's findings~\cite{gerosa_shifting_2021} which showed a recent decrease in \textbf{\emph{own-use}} as motivation for OSS development, our  interview and survey results reveal that personal use remains the dominant motivation for scientific OSS. 
In particular, four out of 11 interviewees (known as long-term contributors) joined the \projectname{} community because their research uses one of the packages in the \projectname{} ecosystem. 
Among all the \numsurveyee{} survey participants (referred to as SP1-80), 53 participants contributed to \projectname{} during their \emph{research/education/hobby-related work} in the science domain. 
Their contributions include fixing bugs or enhancing existing features for \projectname{} as needed for their own research projects. These code changes in turn benefit the community at large.
Most of the cases are self-motivated contributions with some exceptions; in one case, the contribution was by request -- ``\emph{I had found an issue in the software and was asked to contribute a solution to it (if I was able to) by the project maintainers}'' (SP19).

\textbf{Reasons for contributor disengagement.}
Based on the results of our interview and survey, three main issues were identified: lack of motivations, high entry barriers, and conflicts (as shown in Table~\ref{tab:disengage}), which are aligned with Miller's work~\cite{miller2019people}. 
Again, we focus solely on discussing the intriguing findings.

\begin{table}[t]
 \setlength{\tabcolsep}{0pt}
\caption{Reasons for disengagement in \projectname{}.}
\label{tab:disengage}
\footnotesize
\centering
\begin{tabular}{llr} 
\toprule
\textbf{Reasons}                       & \textbf{~Description}                                                                                                                                                                                    & \textbf{\#}  \\ 
\hline
\multirow{3}{*}{ Lack of motivations }     & \begin{tabular}[c]{@{}l@{}}~Focus shifted~(career moves, research/job topics changed)\end{tabular}                                                                                                         & ~53              \\ 
\cline{2-3}
                                    & \begin{tabular}[c]{@{}l@{}}~One-time opportunity (OSS summit ended,~GSoC application failed)\end{tabular}                                                                                       & ~9               \\ 
\cline{2-3}
                                    & ~Project is stable                                                                                                                                                                                   & ~16              \\ 
\hline
\multirow{2}{*}{ High entry  barrier  } & ~For science outsider                                                                                                                                                                                & ~3               \\ 
\cline{2-3}
                                    & \begin{tabular}[c]{@{}l@{}}~For scientists who lack SE background\end{tabular} & ~4               \\ 

\hline
Conflicts                           & ~Conflicts with team members on project plan                                                                                                                                                         & ~1               \\
\bottomrule
\end{tabular}
\vspace{-2em}
\end{table}

A significant portion of survey participants cease their contributions due to \emph{\textbf{lack of motivations}}. In particular, 53 out of \numsurveyee{} survey participants stopped contributing because their \emph{work focus had shifted}, 
such as career changes, graduation, or departure from academia. 
16 of the  \numsurveyee{} participants reported that they are still open to contributing, but the current \emph{project is stable} and already meets their needs.
Additionally, one interviewee and eight survey participants contributed to \projectname{} via \emph{Google Summer of Code (GSoC)}, which is an annual program that pays university students stipends to develop features for various OSS projects~\cite{trainer2014community,gsoc}. 
Only one is currently a core contributor, serving on the management team to handle finance-related responsibilities. 
Two survey participants only contributed one PR each in order to apply for GSoC internship (as part of the application requirement), and they did not continue contributing since their applications were rejected. 
While the other five interns discontinued their participation after the internship, mainly due to shifts in their focus. 
This observation aligns with the observation made by \citet{trainer2014community} who studied community engagement via GSoC and found that it is not a reliable source. 
As mentioned by one interviewee -- from the maintainers' point of view, a major drawback of GSoC PRs is that they tend to be submitted all in a short period of time, causing a sudden surge in workload and with relatively low code quality.

  Another important reason for disengagement is the \emph{\textbf{high entry barrier for both science and SE}}. 
Three participants mentioned the steep learning curve of making a contribution to the project given the domain-specific knowledge underlying the complex code base of its packages. 
Also, four participants explained that without enough SE knowledge, it is difficult to use GitHub and make contributions that meet the requirement defined by the maintenance team. 
Similarly, one core contributor said that it takes at least six months for newcomers to do anything at the expert level.

\begin{figure}[t]
\centering
\includegraphics[width=0.4\textwidth]{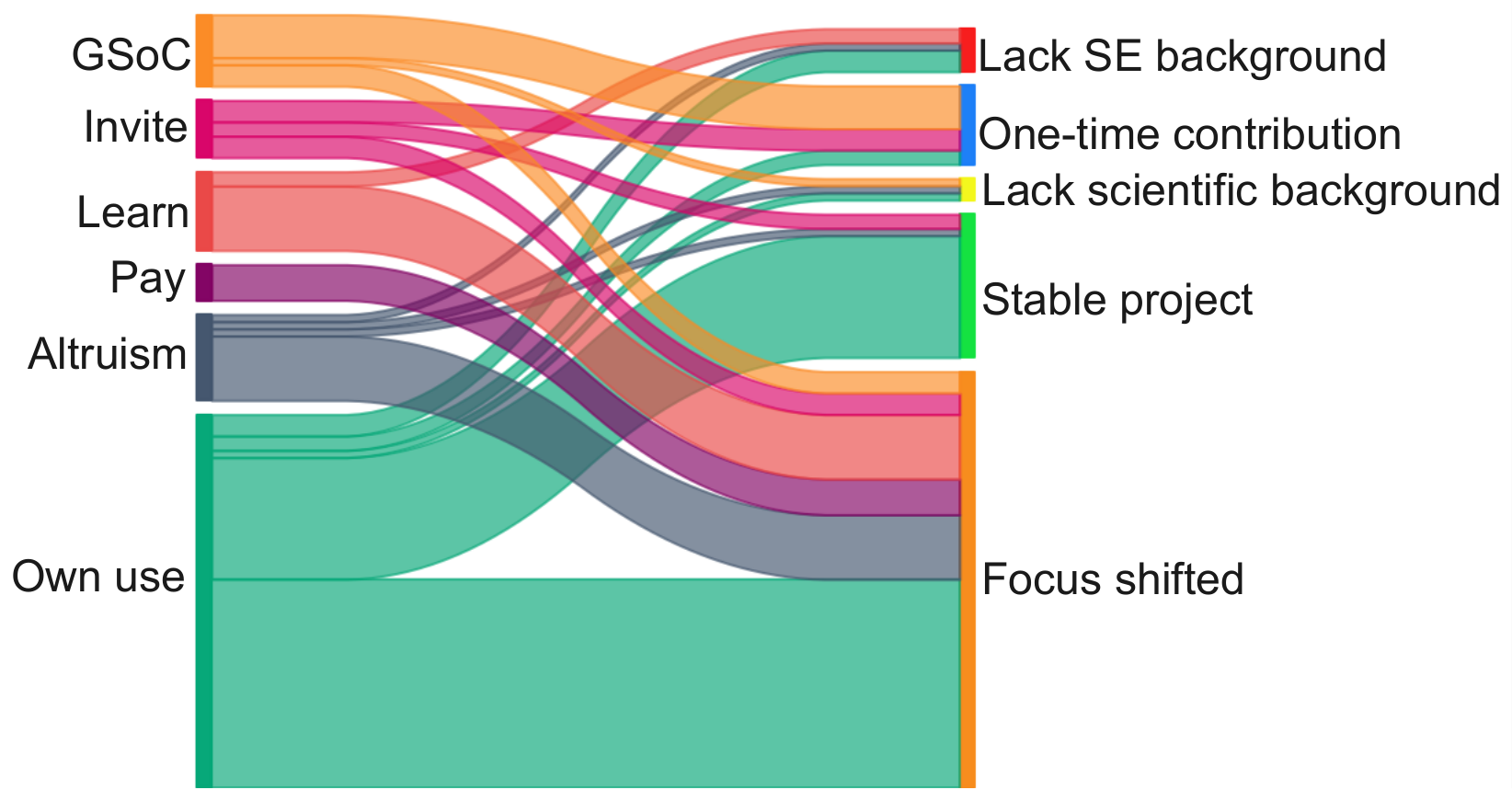}
\vspace{-1em}
\caption{Mapping between motivations and reasons of disengagement from contributors.}
\label{fig:sankey}
\vspace{-1em}
\end{figure}

\textbf{Mapping between motivations and reason for disengagement.}
Lastly, we connect each survey participant's motivations and reasons for disengaging. 
As shown in Fig.~\ref{fig:sankey}, a large portion of the contributors with the motivation of \emph{own-use} stopped after their research focus shifted or the project became good enough for their usage. 
Also, people who made several commits during a short period of time (e.g., GSoC, during research, or OSS summit) stopped contributing right after the end of the event. 
 
\begin{tcolorbox}
[boxsep=2pt,left=2pt,right=2pt,top=2pt,bottom=2pt]
\textbf{Finding 3}: 
In contrast to traditional OSS, where the primary reason for contributing has changed from \emph{own-use} to \emph{altruism} and \emph{learning} in the past decade~\cite{gerosa_shifting_2021}, the \projectname{} community still considers \emph{own-use} as the primary motivation for contributing. 
Regarding contributor turnover, in line with prior research, people stop contributing to scientific OSS mostly because of changes in work focus or due to the high entry barrier for people without domain knowledge and/or professional SE experiences.
\end{tcolorbox}

\subsection{\textbf{Suggestions to Address the Challenges.}}
In this section, we consolidate the viewpoints expressed by both interviewees and survey participants regarding possible strategies to tackle the challenges of sustaining \projectname{} within the context of an open-source community. 
We compile all the suggestions in Tab.\ref{tab:sustain_idea_agg} and  use\science{}to denote that the suggestion is scientific software specific and~\oss{}~to represent that the suggestion is applicable to OSS projects in general, similar to prior work. 
We focus solely on the noteworthy ones for further discussion.

\subsubsection{\textbf{Improving Onboarding \& Retention.}} 
Out of the \numsurveyee{} survey participants, 21 provided insights on how to attract newcomers and retain existing contributors for extended durations. These opinions can be summarized into six categories, as presented in Tab.\ref{tab:sustain_idea_agg}. 
One suggestion that emerged most frequently and holds potential applicability to other OSS communities is to provide ``\emph{\textbf{better support for contribution workflows}}.'' 
One inspiring idea proposed by multiple participants is to 
\emph{better organize ``Good First Issues'' (GFIs)}, 
which is a recommended practice to help onboard newcomers~\cite{xiao2022recommending,GithubGFI}. 
Although it is already adopted by the \projectname{} project, survey participants still find it hard to determine suitable tasks for contribution. 
To mitigate this issue, one core contributor mentioned that the current issue list is too long and hoped to have a ``\emph{better overview of the existing issues for maintainers}'' that links each issue to the code with a summary of the high-level problem. They believe that this would provide a step-by-step guide for solving issues.

\adjustboxset{height=3mm, width=3mm,valign=c, margin=0pt 3pt 0pt 3pt}
\begin{table}[t]
\centering
\caption{Suggestions of improving the sustainability of \projectname{} community to tackle the four challenges}
\vspace{-1em}
\includegraphics[width=0.75\textwidth]{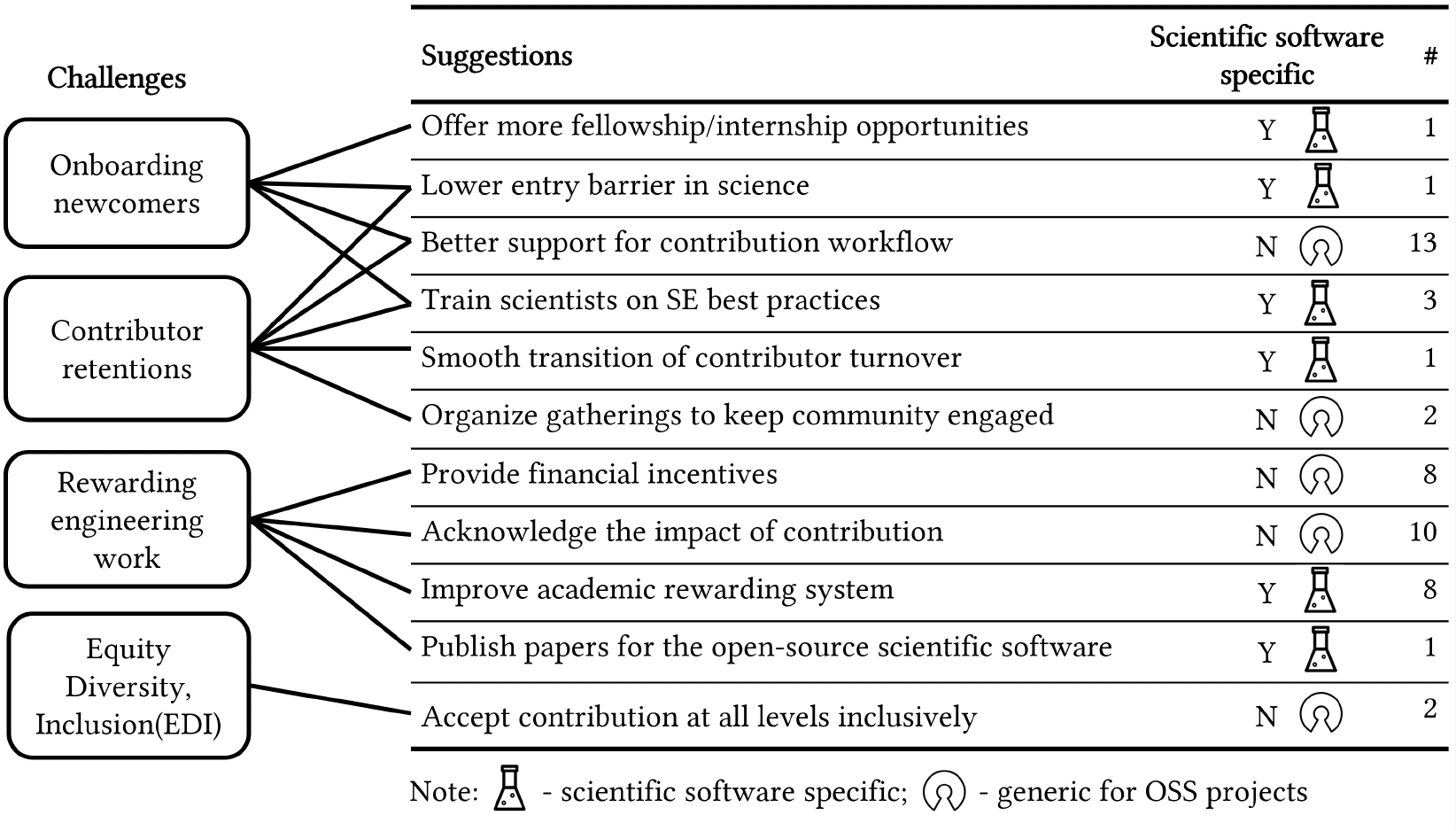}
\label{tab:sustain_idea_agg}
\end{table}

One can argue that a community should not rely too much on graduate students to contribute to the project. However, one survey participant (SP20), who is the founder of one package but left the community due to work shift, said that ``\emph{the best solution is to find a way to pair almost abandoned projects with graduate students who can contribute or even take over the project}.'' 
Given that the shift of research focus is common, a smooth transition can be beneficial for scientific software projects to survive in face of inevitable turnover due to scientific software contributors leaving academia or career change. 
An effective sustainability strategy is essential to proactively address inevitable turnover. For instance, assigning maintenance responsibilities to graduate students and ensuring a seamless transition between maintainers would help alleviate the impact of uncertainties.

\subsubsection{\textbf{Rewarding Engineering Work.}}\label{rewarding}
The professional software engineers in the core team feel that they spend a lot of time helping maintain \projectname{} but their work is neglected because of the \emph{lack of decision-making influence} and \emph{not having enough acknowledgment} from the team. 
When scientists asked if they were aware of such complaints, all of our interviewees acknowledge \emph{the imbalance between too much work and not enough credits}. 
To address this, our study participants offered varying viewpoints.
For instance,%
one intriguing suggestion from four survey participants is to \emph{\textbf{to receive feedback regarding the impact of their work}}. 
As SP13 said that ``\emph{If I had more of a sense for the impact of the bug I discovered, how many people use that function or finding code snippets in other open-source projects which use it, that would help make the intangible benefits more tangible}.''
\szmodifyok{}{In future research, emphasis could be placed on developing methods and metrics to recognize both types of contributions (engineering or scientific) and creating a dashboard to acknowledge contributions of any kind.}

Seven of the survey participants who work in academia (i.e., professors, graduate students) suggested that \emph{\textbf{the academic rewarding systems need improvement to support scientific OSS contribution}}. 
Possible solutions encompass increasing awareness of OSS within scientific communities and ensuring researchers receive due credit for their contributions or maintenance of scientific OSS. Such recognition could serve as a strong motivation for continued contributions.

\szmodifyok{}{It is important to note that the official \projectname{} Team has already taken critical steps in implementing best practices to enhance sustainability within the community~\cite{price2022astropy}. While the effort has been well-received within the community, feedback from non-core members suggests that there is room for further improvement. Future research could compare a larger number of OSS communities who have adopted similar practices and assess the effectiveness and determine whether they have been utilized as intended.}

\begin{tcolorbox}
[boxsep=2pt,left=2pt,right=2pt,top=2pt,bottom=2pt]
\textbf{Finding 4}: In the \projectname{} community, we identified multiple sustainability challenges, some of which are specific to scientific OSS, including the high entry barrier in science, frequent turnover, and undervalued engineering effort in the academic rewarding system. Study participants put forth various solutions to address these challenges, suggesting the development of better tools and rewarding systems, among other ideas.
\end{tcolorbox}

\section{Result: Cross-Project Collaboration (RQ3)}
To understand the collaboration challenges for \projectname{} on the ecosystem level in situ, in this section, we examine how the reference links are used in issues and PRs cross different projects in \projectname{} ecosystem, including the intention of the cross-references and their limitations.

\subsection{Cross-Reference Graphs (CRGs)}

\begin{figure}[t]
\centering
\footnotesize
\includegraphics[width=0.4\textwidth]{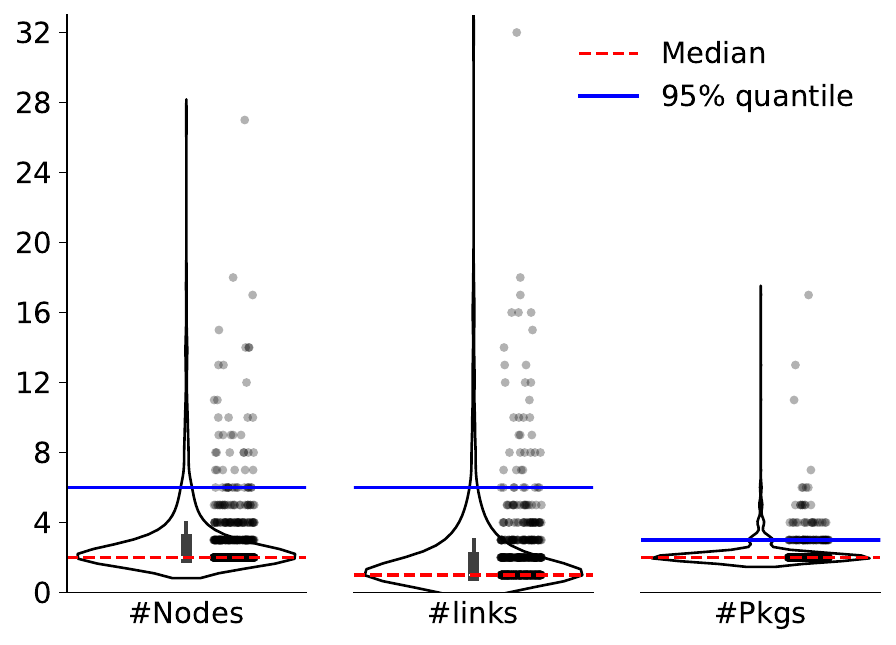}
\caption{Distribution of the number of nodes, links, and packages in one CRG. In total, 697 CRGs.} 
\label{fig:CRG_dist}
\end{figure}

As mentioned in Sec.~\ref{network}, we have collected 1,491 links between pairs of issues/PRs across 43 distinct packages 
contributing to the formation of \numCRGSubgraph~CRGs (denoted as CRG\_1 to CRG\_697). 
These CRGs differ in size measured by node count, ranging from 2 to 27 nodes (mean=2.94, SD=2.18, median=2).
The number of cross-reference links ranges from 2 to 32 per CRG (mean=2.14, SD=2.55, median=1).  
As shown in Fig.~\ref{fig:CRG_dist}, most CRGs involve 2 different packages (mean=2.26, SD=0.99, median=2).
This indicates that although there are cases where multiple issue/PR are discussed together within a common context, the prevailing pattern shows that most cross-project coordination involves two nodes from two different projects connected by one cross-reference link. 
The detailed nodes and links of each CRG could be found in the replication package~\cite{SupMt}.

\subsection{Intention behind Cross-project Collaboration}

\def\themeone{Change coordination due to code dependencies}
\def\themetwo{Knowledge sharing}
\def\themethree{Coordinating shared functionalities}
\def\themefour{Centralized infrastructure management}

Due to the labor-intensive nature of the manual analysis, we conduct the analysis on a sample of the large-size CRGs to infer the collaboration activities as they involve multiple projects. 
In particular, we selected the 95\% quantile from all \numCRGSubgraph~CRGs, resulting in 34 CRGs with seven or more nodes and 375 edges. 

\begin{figure}[t]
\centering
\footnotesize
\includegraphics[width=0.75\textwidth]{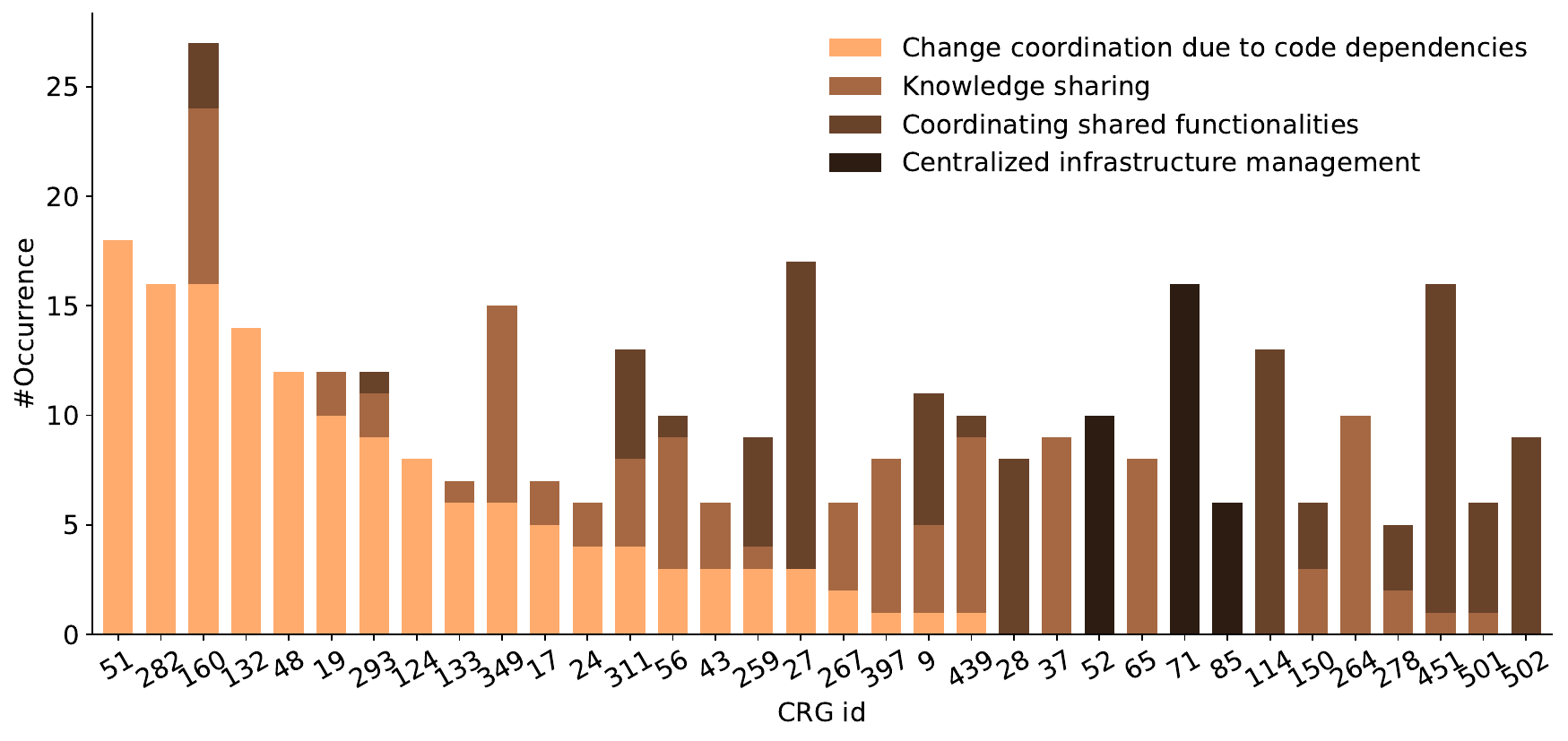}
\caption{Distribution of CRG labeling results. Each bin represents one CRG. The different colors indicate the different themes identified in each edge within the CRG. The bins are arranged in descending order based on the number of edges in the ``change coordination due to code dependencies'' category.} 
\label{fig:CRG_label}
\end{figure}

Following the qualitative analysis procedure (described in Sec.~\ref{network}), 
we identified themes concerning the intention behind the cross-project collaboration. 
During the process,  we excluded 9 edges that were mistakenly created by the developers in 3 CRGs. 
Consequently, we identified four themes from 366 edges, including (1) \themeone, 
(2) \themetwo, 
(3) \themethree, and 
(4) \themefour. 
We presented the distribution of the four themes in 
Table~\ref{tab:CRG_dist}. 
Most CRGs have more than one theme identified. 
In Fig.~\ref{fig:CRG_label}, we depict the composition of link types in each CRG.

\begin{table}[t]
\begin{threeparttable}[b]
 \setlength{\tabcolsep}{0pt}
\caption{Distribution of CRG labeling results}

\label{tab:CRG_dist}
\footnotesize
\centering
\setlength\extrarowheight{0pt}
\begin{tabular}{lcr} 
\toprule
\textbf{Collaboration intention }
& \textbf{~CRG appearance count}
& \textbf{~Links count/ratio}  \\ 
\hline
Change coordination due to code dependencies
& \begin{tabular}[c]{@{}l@{}}~ 21\end{tabular}
& 145 (39.63\%)
\\
\hline
Knowledge sharing
&~\begin{tabular}[c]{@{}l@{}}22\end{tabular} 
&~97 (26.5\%)
\\
\hline
Coordinating shared functionalities
&~\begin{tabular}[c]{@{}l@{}}15\end{tabular} 
&~92 (25.14\%)
\\
\hline
Centralized infrastructure management                           
& \begin{tabular}[c]{@{}l@{}}~3\end{tabular}
& ~32 (8.74\%)
\\
\bottomrule
\end{tabular}
\end{threeparttable}
\end{table}

\subsubsection{\textbf{\themeone}.}\label{change-impact}
This type of cross-project collaboration arises from the interconnection of code dependencies between upstream and downstream packages, constituting 39.63\% of the links. 
To gain an overview of the code dependency within the \projectname{} ecosystem, 
we parse the PyPI installation requirement\footnote{Python packaging metadata for distribution requirement:\url{https://packaging.python.org/en/latest/specifications/core-metadata/\#requires-dist-multiple-use}} among the 51 packages to identify their direct upstream. 
The results show that 46 out of the 50 \emph{coordinated} or \emph{affiliated packages} directly depend on the \emph{core package}, with many packages being also interdependent with each other. 
This phenomenon is akin to previous research exploring the interdependence of projects within a software ecosystem. 
Because of the interdependent code relationships of the packages in the ecosystem, 
it is not surprising that resolving dependency-related issues such as version syncing, refactoring as reactions to API changes~\cite{hora2015developers,robbes2012developers}, 
breaking changes~\cite{Xavier2017Historical,bogart2016break}, 
and cross-project bug fixing~\cite{ma_how_2017,ma2020impact}, etc (as described in Sec.~\ref{related}).

\subsubsection{\textbf{\themetwo}.} \label{knowledge-sharing}
Cross-project links are also used to facilitate knowledge sharing and improve collaborative discussions. 
This theme accounts for 26.5\% of the categorized links across 22 CRGs. 
Considering that all packages within the \projectname{} ecosystem fall under the astronomy domain, it is unsurprising that solutions found for one package could serve as inspiration for others. 
People from different communities often help each other 
to better understand the rationale behind the code changes or feature implementation. 
With an increased number of linked issue/PRs and the prolonged duration of these discussions, individuals often experience information overload, 
making it difficult to keep track of the conversation. 
For example, during the discussion of fixing a bug (CRG\_37), one long-term contributor pointed out another issue that proposed exactly that 7 years). 
This problem could also hinder the contributors' ability to monitor changes and comprehend modifications made by others in the future. 
Relevant concern was also mentioned by our interviewees when describing their management workload of one package on GitHub:
``\emph{... issues and PRs have been a really important way for us to keep track of everything that's going on with the project. But I think we're hitting the limits of what that kind of interface can do for collecting the relevant information...we want people to be able to know like where are the places that actually that we need help on, and so I wonder if there's a better way of consolidating that information than to have 1000 open individual issues.}''

\subsubsection{\textbf{\themethree.}} \label{common-feature}
Contributors from multiple communities also engage in communication related to new features, refactoring, and/or migrating existing features. 
Among the 34 CRGs we examined, 15 exemplify situations in which projects collaborated to enhance features to achieve better usability, leading to mutual benefits across multiple projects. 
Overall, we recognized 92 links falling into this category, making up 25.14\%. 
A typical example is in CRG\_28, where the conversation originated from a 
feature request of ``the ability to select existing observatories for EarthLocation'' (\emph{astropy/astropy}, Issue\#3813). 
The feature was initially implemented in a \emph{coordinated package (astroplan)}, and 
ported to the \emph{core project astropy/astropy}
to benefit a broader audience. 
Eventually, after one year, contributors from both projects collaborated to eliminate redundant code 
and incorporate additional test cases, ensuring the feature's overall quality.

It is important to note that not all initiated conversations led to a resolution  that met the expectations of every stakeholder involved. 
In many cases, \emph{\textbf{the presence of duplicate feature implementation and fragmented development efforts is inevitable.}} 
For instance, in CRG\_9, the contributor of \emph{gammapy} expressed their demand for a feature 
regarding location data access and
acknowledge that similar features already exist in two other projects.  
Due to an impending deadline for a workshop and stalled progress of a similar feature in another project, the contributor decided to implement their own version and 
hoped that the three packages ``can be merged into one glorious regions package at some point.''

Furthermore, \emph{\textbf{divergent requirements from various communities can result in conflicts and sub-optimal solutions}}.
For example, in CRG\_451, in 2018, contributors from three projects (\repoa, \repob, and \repoc) 
found that they share 
similar features with different implementations under different licenses 
that could be combined and unified, preventing duplicated efforts in the future. 
Given that Project \repoa \, would like to ``split and optimize the libraries to make the code base lighter,''  
the maintainer of \repoa~ created one PR to `donate' one \emph{NEOs} module to Project \repoc~to make it even more usable (\emph{astroquery}, PR\#1325),
and another PR to merge the \emph{Dastcom5} module to Project \repob~(\emph{sbpy}, PR\#115).
However, neither PR was merged in the end. The first PR got rejected by \repoc~ maintainers because the code ``was a subset of already existing functionality without providing much functionality complementary.'' 
The maintainer in \repoa~ then decided to remove the \emph{NEOs} module entirely 
since it had become burdensome to maintain and \repoc~provides a potentially better alternative. 
They also noted the downside of this move is losing the authorship of the code.
The second PR was ignored for over four years and is still pending due to a long period of inactivity in \repob~and \szmodifyok{}{thus the effort of creating the PR was wasted.}

\subsubsection{\textbf{\themefour.}} \label{infra-mgmt}
Last but not least, we identified 32 links (8.74\%) from three CRGs (ID\_52, 71, 85) that represent the cross-project coordination related to the shared infrastructure supporting the community's software development activities. 
As described in Sec.\ref{astropy-intro}, the \projectname{} Team supports the broader ecosystem by providing pre-configured infrastructure packages that facilitate the maintenance effort for each package. 
The maintenance of the infrastructure is operated in a centralized way. 
Similar practices have been applied in other OSS ecosystems such as Numpy~\cite{harris2020array}. 
The cross-reference links were created by the official \projectname{} team maintainers to broadcast code changes to other packages using the infrastructure. 
For example, in CRG\_52, 11 projects participated in the conversation about updating the code coverage tool that is shared by many packages~(\emph{astropy/astropy}, PR\#12245).
Further, the maintainer of the core package automatically opened 11 PRs using \emph{batchpr}~\cite{batchpr} under each repository to patch the affected code blocks. 
Such practice mitigates the maintenance burden of the affiliated packages' management team. 
\szmodifyok{}{Future research could explore the extent of prevalence of this centralized infrastructure management approach and understand its impact on reducing the maintenance burden and potentially enhancing sustainability at the community level.}

\begin{tcolorbox}[boxsep=2pt,left=2pt,right=2pt,top=2pt,bottom=2pt]
\textbf{Finding 5}: We observe four types cross-project collaboration, including (1) \themeone, (2) \themetwo, (3) \themethree, and (4) \themefour to support issue discussions, as well as helping others better understand the scientific theory behind the code design. 
We also identified inefficiencies during collaboration such as duplicate code, fragmented implementation, wasted effort, and lack of awareness within the ecosystem.
\end{tcolorbox}

\section{Implication}

Our study reveals that while  \projectname{} 
serves as a relatively prosperous instance of a scientific OSS ecosystem, 
it still confronts challenges that hinder the sustainability of the software, the community, and the ecosystem. Here, we discuss implications and recommendations for different stakeholders to meet those challenges.

\textbf{Implication for scientific OSS practitioners.} 
While scientific OSS faces sustainability challenges similar to generic OSS communities, our investigation revealed that the entry barrier is elevated. This is attributed to the requisite domain-specific knowledge during development, necessitating both software engineering expertise and a scientific background for meaningful contributions to the project. 
Therefore, lowering the entry barrier for contribution by providing detailed documentation that explains software-related information such as code design rationales can enhance the understanding of relevant concepts and facilitate informed contributions. Tutorials, workshops, and effective tooling support are also important in expanding the user base and turning existing users into contributors. 
Listing the required expertise, including both scientific and software-related aspects, into the ``Good First Issues'' 
can help potential contributors better select and succeed in the tasks. 
Furthermore, within scientific OSS communities, contributions related to engineering are often undervalued, primarily due to the academic evaluation system. To ensure the sustainability of both the community and the software, it is crucial to design methods to recognize and acknowledge contributions to the code base as a means of retaining existing contributors.

\textbf{Implication for SE researchers and tool builders.}
How to substantially reduce the workload for both scientists and software engineers remains a central concern for designing tools serving the scientific OSS community. 
Future studies could explore approaches to categorize issues and offer guidance that includes the essential scientific theory behind the issue and the necessary programming knowledge. 
Tools to break down the sub-tasks necessary to resolve the related issues would also be beneficial in emulating a step-by-step onboarding process for newcomers. 

Moreover, our manual analysis of cross-project collaboration reveals a considerable amount of valuable information embedded and also scattered in the different issue/PR discussions across multiple projects. 
Navigating relevant information within the ecosystem becomes increasingly challenging and overwhelming when tasked with a specific objective. 
Subsequent research endeavors could explore the tooling support required to generate a better overview and effectively organize issues/PRs discussions beyond a single thread~\cite{summit_cscw2023,kumar2023summarize}, 
potentially involving extracting valuable information and generating project documentation, aiming to enhance knowledge sharing while also ensuring that important information is not lost or scattered without appropriate management. 

Moreover, future studies could delve into methods for inferring the intention behind cross-referenced links within and between projects, aiming to enhance the support provided for the knowledge extraction task mentioned earlier.

To appropriately acknowledge contributors and encourage long-term commitment, as highlighted by our survey participants, it is essential to demonstrate the impact of their contributions both quantitatively and qualitatively. Therefore, future efforts could concentrate on developing improved metrics and detection tools that showcase the contribution's impact beyond merely counting the downloads and/or citations of scientific open-source software.

\textbf{Implication for funding agencies/research institutions.}
Scientists hold different mindsets when it comes to task prioritization during software development. 
It is not surprising that scientists lack the motivation to invest effort into code quality, given that academic reputation is mainly based on scientific contribution~\cite{howison_incentives_2013}. 
This could lead to rejection and abandoned PRs during the code review process, and discourages scientists from making continuous contributions.
It is important to provide institutional support to acknowledge the contribution to engineering work alongside the scientific ones, 
such as creating career paths for scientists in research SE and providing incentives for them to invest more effort in improving software quality. 
Meanwhile, allocating long-term financial support for maintaining the necessary infrastructure for scientific OSS is also vital.

\section{Conclusion}
In this study, we study the sustainability challenges of a scientific OSS ecosystem. 
Specifically, we investigate the problem from both the interdisciplinary collaboration, the communities and the ecosystem aspects. 
Our results show that there are many unique properties in the scientific domain that are different from the traditional OSS. 
We also propose future research and tooling directions to address the challenges.

\bibliographystyle{ACM-Reference-Format}
\bibliography{references}

\end{document}